\documentstyle[11pt,aaspp4]{article}

\lefthead{Korchagin, Kikuchi}
\righthead{}
\newcommand{\be}{\begin{equation}}
\newcommand{\ee}{\end{equation}}
\begin{document}

\title{Global Spiral Modes in NGC 1566: Observations and Theory}
\author{V. Korchagin$^1$}
\affil{Institute of Physics, Stachki 194, Rostov-on-Don, Russia\\
Email: vik@rsuss1.rnd.runnet.ru}
\altaffiltext{1}{National Astronomical Observatory, Mitaka,
Tokyo 181--8588, Japan}
\author{N. Kikuchi}
\affil{Center for Computational Physics, University of Tsukuba, Tsukuba, 
Ibaraki 305--8577, Japan\\
Email: kikuchi@rccp.tsukuba.ac.jp}
\author{S. M. Miyama}
\affil{National Astronomical Observatory, Mitaka, Tokyo 181--8588, Japan\\
Email: miyama@yso.mtk.nao.ac.jp}
\author{N. Orlova}
\affil{Institute of Physics, Stachki 194, Rostov-on-Don, Russia\\
Email: nata@iphys.rnd.runnet.ru}
\and
\author{B.A. Peterson$^1$}
\affil{Mount Stromlo and Siding Spring Observatories,
RSAA, ANU, Weston Creek, ACT, 2601, Australia\\
Email: peterson@mso.anu.edu.au}
\begin{abstract}
We present an observational and theoretical study of the spiral structure in
galaxy NGC 1566.
A digitized image of NGC 1566 in $I$-band, obtained with the 
Australian National University 30in telescope, was used for 
measurements of the radial dependence of amplitude variations in the
spiral arms.
The azimuthal variations of the surface brightness in the $I$-band 
are about $\pm 7$ \% at $50^{\prime \prime}$ radius and increase
up to $57$ \% at $100^{\prime \prime}$.

We use the known velocity dispersion in the disk of NGC 1566, together with
its rotation curve, to construct linear and 2D nonlinear
simulations which are then compared with observations.
The linear stability analysis of the disk of NGC 1566, made under
the assumption that the ratio of the vertical and the radial velocity
dispersions is equal to 0.6, shows that the disk is stable with respect to
spiral perturbations.
We confirm this conclusion with 2D hydrodynamical simulations.
The disk of NGC 1566 is unstable if the
ratio of the vertical and the radial velocity dispersions $c_z/c_r$ is
of order $0.8 - 1.0$. Under this assumption, a two-armed spiral 
constitutes the most unstable global mode in the disk of NGC 1566.

The 2D hydrodynamic simulations of the unstable disk seeded with 
random perturbations show the exponential growth of two unstable modes,
$m=2$, and $m=3$.
The growth rates of the global modes seen in the nonlinear simulations
are in good agreement with the results of the linear modal analysis.
The $m=3$ global mode is less important, however, in comparison with the main
$m=2$ global mode, and the overall evolution of the perturbations is
determined by the two-armed spiral mode, which saturates at a level
$\log_{10} A_2 \approx -0.5$.
The theoretical surface amplitude and the velocity residual variations
across the spiral arms calculated during the nonlinear phase of instability
are in qualitative agreement with the observations. 

The spiral arms found in the linear and nonlinear simulations are
considerably shorter than those observed in the disk of NGC 1566.
The nonlinear phase of instability is characterized by the transport of
angular momentum towards the disk center, making the surface density 
distribution quite steep in comparison to the observations.
We argue therefore, that the surface density distribution in the disk of
the galaxy NGC 1566 was different in the past, when spiral structure in
NGC 1566 was linearly growing.
\end{abstract}
%
%
      \section{INTRODUCTION}

There is a general consensus that the regular spiral arms observed in 
disk galaxies are a manifestation of the global density waves 
propagating in galactic disks (Binney \& Tremaine 1987).
In the approach advocated for many years by C. C. Lin and his collaborators
(e.g. Bertin et al. 1989), such global waves arise from the gravitational
instability of exponentially growing internal modes which
are stabilized at some ``quasi-steady'' level due to nonlinear 
effects. Another explanation assumes that the global
spiral arms arise from swing-amplified chaotic fluctuations
originating from N-particle noise (James \& Sellwood 1978),
or from the gravitational drag of the co-orbiting mass clumps within the
disk (Toomre 1990).
A long-standing controversy over the physical mechanisms
responsible for generation of the global density waves 
has recently been resolved by Shu et al (1999).
Those authors showed that global structures
arising from the internal normal modes of a gravitating galactic disk,
and the spiral patterns driven by external
tidal influences, are both manifestations of the fundamental underlying process 
of modal selection and growth.

The discovery of the protoplanetary disks around T Tauri stars
initiated a new wave of interest in global
instabilities in self-gravitating disks.
A number of authors have developed various techniques to
study linear density perturbations in gaseous and the
collisionless self-gravitating disks 
(Adams, Ruden, \& Shu 1989; Noh, Vishniac, \& Cochran 1991;
Laughlin \& R\'o\.zyczka 1996; Vauterin \& Dejonghe 1996;
Kikuchi, Korchagin, \& Miyama 1997).
Linear stability analysis methods for self-gravitating disks
are well complemented by direct two-dimensional simulations
(e.g., Tomley et al 1994; Miyama et al. 1994;
Nelson et al. 1998).
Zhang (1996), and independently Laughlin, Korchagin, \& Adams (1997, 1998) 
have examined nonlinear self-interaction of a single strong global mode.
They found that nonlinear self-interaction is
responsible for the nonlinear saturation of the mode, and concluded that
the saturation mechanism provides a key for understanding
the long-term behavior of spiral density waves.

An ultimate aim of any theory of spiral structure is the construction
of models of particular galaxies on the basis of the available
observations.
A number of authors have undertaken such comparisons
(e.g., Lin, Yuan, \& Shu 1969;
Roberts, Roberts, \& Shu 1975; Mishurov et al 1979;
Elmegreen \& Elmegreen 1990).
The procedure in these studies 
has relied on empirical estimates of the positions of the
corotation and Lindblad resonances using ``optical tracers'' in the
galactic images.
Local dispersion relations were then used to calculate the spiral
response under additional assumptions regarding the radial behavior of the
stability parameter $Q$ derived by Toomre (1964).
These two foregoing assumptions,
and especially the stipulated choice of 
Toomre's $Q$-parameter, considerably reduce the predictive power of the
spiral density wave theory.

Long-slit spectroscopic observations of spiral galaxies can provide a stronger
basis for the comparison of density wave theory with
observations.
Knowledge of the $z$-velocity dispersion in a galactic disk
provides a quantitative estimate of the surface density distribution
under the realistic assumption that disks are self-gravitating in
the $z$-direction (Bottema 1992, hereafter RB).
The velocity dispersion, together with the known rotation curve
and the surface density distribution, uniquely determine the
axisymmetric background state of the galactic disk, and can be used for
linear and nonlinear analyses aimed at the modeling of individual spiral
galaxies.

The aim of our paper is to model the spiral structure in the nearby
grand design spiral galaxy NGC 1566. 
The galaxy is apparently isolated, and it has been the
subject of many photometric and spectroscopic studies.
In particular, the rotation curve and the radial profile of the stellar
velocity dispersion have been accurately measured (RB).  NGC 1566 
represents one of the best candidates for the comparison of spiral density
wave theory with observations.

In contrast to previous comparisons between theory and observations,
our study models the spiral arms in the disk of
NGC 1566 from ``scratch'', without making any additional
assumptions with regards to the positions of the principal resonances
in its disk or the behavior of Q-parameter.
Basing our work on both our own, as well as previously published
observations,
we undertake the linear global modal analysis of this
system, and extend the analysis  
of the spiral structure into the nonlinear regime using 2D numerical
simulations. We then compare the results of our linear,
and nonlinear simulations
with the observed properties of the spiral arms in NGC 1566.
The purpose of this comparison is to determine the extent to which
the internal global modes emerging from random
perturbations imposed on the gravitating disk of NGC 1566
can be responsible for the observed properties of its spiral
arms.

In Section 2 we present 
the optical maps in $B$- and $I$-bands for NGC 1566, along with
the results of our measurements of the
amplitude variations in the spiral arms.
In Section 3, we use the available observational data to determine
the background equilibrium properties in the disk of NGC 1566. 
In sections 4 and 5 we describe the basic equations and our
computational methods. Section 6 gives the results of the
linear global modal analysis. Section 7 describes the results
of the nonlinear 2D simulations and discusses our comparison 
of the theoretical results with observations.  


	\section{OBSERVATIONS}

\subsection{Photographic Observations and Data Reduction}

The data consists of 13 B-band images and 9 I-band images, 
each with an exposure time of 300 seconds. The images
were obtained with the 30in telescope at Mt. Stromlo Observatory
(35.32 degrees south latitude, 149.00 degrees east longitude, 
elevation 770 meters) by the RAPT observers on the night of
12 November 1999. The detector used was a liquid
nitrogen cooled, thinned SITe CCD with 1024x1024 square 24 micron 
(0.36 arc-sec) pixels. A correction for a bad column
was made by interpolating the intensities of the adjacent
columns. Each image was processed by subtracting a bias image, 
dividing by a flat field image for the appropriate band,
spatially filtering to remove cosmic rays, and subtracting
the mean sky. The images were then aligned, and those for each 
band were combined by taking the median value in each pixel.
Observations were made of Landholt CCD standards on the same night. 

\subsection{Inner Spiral Structure in NGC 1566}

The morphology of NGC 1566 has been studied observationally in different
bands by number of authors (Vaucouleurs 1973; Hackwell \& Schweizer 1983;
Pence, Taylor, \& Atherton 1990).
The most prominent morphological feature of NGC 1566 is a near-perfect
two-armed spiral pattern observed in its inner parts
($3.5^{\prime} \times 3.5^{\prime}$).
Pence et al. (1990) report that the faint outer spiral arms of NGC 1566
are $7.5^{\prime} \times 7.0^{\prime}$ in size.
The broader outer spirals are a continuation of the bright inner spiral
pattern, and form a pseudoring at the periphery of the galaxy.
Infrared photometry (Hackwell \& Schweizer 1983) indicates the
presence of a short bar structure.

Figure 1 shows the superimposed $B$- and $I$-band images
of the inner spiral structure of NGC 1566.
The blue HII regions are arrayed along the inside edge of the inner
spiral arms. Outside of the narrow ridge of blue light, is a broad yellow arm
of older stars which decreases its brightness while moving radially
outward. 

The digitized image in $I$-band (Figure 2) was used for the measurements of the
radial dependence of surface brightness variations in spiral arms.
Figure 3 shows the azimuthal profiles of logarithm
of the number of counts averaged
in six rings at radii $20^{\prime \prime}$, $30^{\prime \prime}$,
$50^{\prime \prime}$, $60^{\prime \prime}$, $80^{\prime \prime}$ 
and $100^{\prime \prime}$ with thickness $5.4^{\prime \prime}$ each.
This value, proportional to the surface brightness, 
gives the azimuthal variations of the surface
brightness associated with the spiral arms.  

The amplitude variations 
measured from the mean surface brightness in a given ring increase with radius. 
The result concurs with observations of Hackwell \& Schweizer (1983),
and with the analysis of the optical image of NGC 1566 made by Elmegreen
\& Elmegreen (1990) who also found an increase of the amplitude
variations in spiral arms with radius.  The surface brightness variations
are about 7 percent at radius $50^{\prime \prime}$,
and increase up to 57 percent at radius $100^{\prime \prime}$. 
The overall spiral amplitude variations in I-band agree with observational
data of Hackwell \& Schweizer (1983) who found in the infrared
spiral amplitude variations about $\pm 33$ percent at $50^{\prime \prime}$ 
radius.


\section{EQUILIBRIUM MODEL}

Observational data do not provide a unique set of background
equilibrium functions for NGC 1566. We therefore discuss the
stability properties of a family of disks allowed within the
observational error bars. 
For numerical purposes, it is convenient to use units in which
the gravitational constant $G$ is equal to unity, the unit of mass is equal to
$10^{10} M_{\odot}$, and unit of length is 2 kpc.
The velocity and time units are then equal to 149.1 km/sec and $1.34 \times 10^7$ years.
We use these units throughout the paper except specially mentioned
cases. The units we use are different from the galactic units in common use,
e.g. those defined by Mihalas and Routly (1968) ( 1km/sec, 1 kpc,
$2.32\times10^5 M_{\odot}$). Our units are more advantageous
in the numerical simulations. With these units, the typical
dimensionless rotational velocity of the disk and its total mass
are of order of unity which is essential in choice of the
timestep in the 2D integration. Similar units are
used in the 2D numerical simulations of gravitating protoplanetary
disks (e.g. Laughlin \& R\'o\.zyczka 1996).

{\it Surface Density Distribution.} 
In the $H$-band, the surface brightness distribution in NGC 1566 is
represented by an exponential law with a radial scalelength $h_{SB} =
15.5^{\prime \prime}$, corresponding to 1.3 kpc for the adopted distance
17.4 Mpc.
The $z$-component of the stellar velocity dispersion in NGC 1566 is well
fitted by an exponential law with a radial scalelength approximately
twice as large as the scalelength of the surface brightness
distribution (RB), or with the $z$-component of the velocity dispersion
proportional to the square root from disk's surface brightness. 
Exponential distributions of the velocity dispersion with its value
proportional to the square root of the surface brightness 
were found also in other disk galaxies (Bottema 1993).
This fact is consistent with the assumption that the galactic disks can
be represented by the locally isothermal, self-gravitating stellar
sheets with the surface density $\sigma(r)$, and the vertical velocity
dispersion $c_z$  related as:
\be
\sigma(r) = c_z^2/\pi G z_0
\label{eq1}
\ee
Here $G$ is the gravitational constant, and $z_0$ is the effective
thickness of the disk.
Van der Kruit \& Searle (1981) in a study of edge-on disk galaxies found
that, to rather high accuracy, the parameter $z_0$ does not depend on
the radius.
We use this assumption in our modeling.

RB found that equation (1) provides a good fit to the observational
profile of the velocity dispersion with the constant effective thickness
of the  disk $z_0 = 0.7$ kpc, and with the mass-to-light ratio $(M/L)_H
= 0.45 \pm 0.15$. 

The value of the central velocity dispersion can be considered
separately. Bottema finds that the core
of NGC 1566 has a constant velocity dispersion
$115 \pm 10$ km/sec. This value does not fit the
overall exponential distribution of the velocity dispersion
within the NGC 1566 disk, which has a known scalelength $31^{\prime \prime}$.
Furthermore, the 115 km/sec central velocity dispersion does not match
the $80 \pm 15$ km/sec value measured at radius $22^{\prime \prime}$
by van der Kruit \& Freeman (1984) and reconfirmed by RB.
Bottema thus assumed that the anomalous velocity dispersion in the core
is related to the small central bulge.
In our model, we choose the central value of the velocity 
dispersion to be 155 km/sec, which gives a satisfactory fit
to the observed radial distribution of the velocity
dispersion in the NGC 1566 disk.

The exponentially decreasing stellar velocity dispersion is about
3 km/sec at radius $120^{\prime \prime}$, corresponding to 10 kpc for
the adopted distance.
We therefore choose a 10 kpc outer radius for the disk of NGC
1566.  

Once the $z$-component of the stellar velocity dispersion is fixed, 
equation (1) provides the surface density distribution of
the NGC 1566 disk.
For the numerical reasons, we assume that the surface density distribution 
goes to zero at the outer boundary of the disk, 
and use a surface density distribution of the form:
\be
\sigma(r) = \sigma_0 \exp(-r/h_{\sigma})[1-(r/R_{out})^2]^5 {\mbox ,}
\label{eq2}
\ee
This distribution incorporates
a dimensionless factor $[1-(r/R_{out})^2]^5$ which causes
the surface density at the outer edge of the disk to vanish in a smooth
way. The functional form (2) provides a good fit to the observed exponential
distribution of the surface brightness in the disks of galaxies.
Note, however, that the global stability properties are insensitive
to the particular choice of the functional forms of the equilibrium
functions.

The corresponding $z$-component of stellar velocity dispersion is thus given
by the equation:
\be
c_z(r) = c_{z0} \exp(-r/2h_{\sigma})[1-(r/R_{out})^2]^{2.5}
\label{eq3}
\ee
The assumed value 155 km/sec of the velocity dispersion at the center
of the disk yields a central surface density of
$\sigma_0 = 2.44 \times 10^9 M_{\odot}/{\rm kpc}^2$, and a total  disk mass
equal to $1.79 \times 10^{10} M_{\odot}$.

{\it Rotation Curve.} With an inclination of $28^{\circ}$, 
NGC 1566 is not
well suited for an accurate determination of the rotation curve.
The medium curve on Figure 4 indicated by pluses shows the observed
rotational velocity of NGC 1566 in units 149.1 km/sec taken from RB.
Errors indicated by the upper and lower dotted curves were
also taken from RB, and arise from an
uncertain knowledge of the inclination of NGC 1566, which was estimated by RB
to be $\pm 5^{\circ}$. 

In modeling NGC 1566's dynamical properties, we have adopted the rotation
curve given by the equation:
\be
v_0(r) = {V_1 r \over (r^2 + R_1^2)^{3/4}}
       + {V_2 r \over (r^2 + R_2^2)^{3/4}}
\label{eq4}
\ee
This rotation curve exhibits the basic features of
galactic rotation.
It goes to zero for small $r$, is ``flat'' in the outer regions of the
disk, and presents a Keplerian decline in the regions beyond the
galactic halo. The rotation law (4) differs from the model rotation curve
used by Bertin et al (1989). We found, however, that the model
curve (4) better reproduces observed rotation in NGC 1566.
Figure 4 superimposes this rotation curve
on the observational data.
The parameters $V_1, V_2, R_1,R_2$, determining the ``low'', ``medium'',
and ``upper'' rotation curves are given in Table 1.

{\it Radial Velocity Dispersion.} 
Theoretical arguments based on the results of N-body
experiments (Villumsen 1985) suggest that the ratio of vertical and azimuthal
velocity dispersions is constant throughout the galactic disks, and 
maintains a value of roughly 0.6.
RB found this value to be adequate in his analysis of the stability
properties of NGC 1566.
We therefore adopt the ratio $c_z/c_r= 0.6$ as a ``standard'', while
also considering higher values $c_z/c_r= 0.8$, and $c_z/c_r= 1.0$.

Radial velocity dispersion together with an epicyclic frequency
and a surface density determines the Toomre $Q$ parameter which is an 
important measure of the stability of a gravitating
disk. For a stellar disk, the $Q$ parameter is given by
\be
Q = {c_r(r) \kappa(r) \over 3.36 \sigma(r)}
\label{eq5}
\ee
Here $\kappa(r)$ is an epicyclic frequency and $c_r$ is the radial
velocity dispersion in the disk.

{\it Fluid Approximation.}
In our analysis of the stability properties of NGC 1566, we use the
hydrodynamic approximation, modeling the stellar disk as a fluid with a
polytropic equation of state with the polytropic index $\gamma$:
\be
P_s = K_s \sigma(r)^{\gamma} 
\label{eq6}
\ee
Here $P_s$ is the vertically integrated pressure and $K_s$ is the
polytropic constant.
The radial velocity dispersion is then related to the surface density by:
\be
c_r = (\gamma K_s \sigma(r)^{\gamma-1})^{1/2}
\label{eq7}
\ee
In our simulations we choose the value of polytropic constant
$\gamma = 2.0$ which naturally produces the empirical ``square root''
proportionality between the velocity dispersion and the surface density
found in galactic disks.

There are some additional arguments justifying the hydrodynamic
approximation in the description of stellar disks. In the papers of
Marochnik (1966), Hunter (1979), and Sygnet, Pellat, \& Tagger (1987)
it was shown that in some particular models the behavior of
perturbations in the collisionless disks can be described by introducing
an isotropic pressure with  polytropic constant  $\gamma =2$, or alternately,
a pressure tensor.
Using the hydrodynamic approximation, Kikuchi et al. (1997) made a direct
comparison of the global stability properties of gravitating disks with
solutions of the collisionless Boltzmann equation obtained by
Vauterin \& Dejonghe (1996).
Kikuchi et al. found good qualitative, and in most cases good quantitative
agreement between the results obtained in the hydrodynamic
approximation, and those found by the direct solution of Boltzmann equation.

{\it Halo Potential.}  The equilibrium rotation of the disk $v_0(r)$ is
balanced by the external gravitational potential of the rigid halo
potential $\Psi_H$, the self-gravity of the disk $\Psi$, and the radial
pressure gradient:
\be
{v_0^2 \over r}
   = {1 \over \sigma} {d P_s \over d r}
   + {d \over dr} \Big( \Psi + \Psi_H \Big )
\label{eq8}
\ee
Since the surface density distribution and the equilibrium rotation curve are
fixed by equations (2) and (4), the self-gravity of the disk can be
found by numerically solving Poisson's equation, and equation
(8) can be used to calculate the gradient of the halo potential $d
\Psi _H /dr$. The total mass of a spherically symmetric rigid halo
within a radius $r$ is thus determined by the expression:
\be
M_H(r) = r^2 {d \Psi _H(r) \over dr}
\label{eq9}
\ee
 
\section{ BASIC EQUATIONS}

Galactic disks are the multi-component systems containing collisionless
stars and gas in different phases.
The common approximation in studying the spiral structure of galaxies is
an assumption that the galactic disks can be considered as a
one-component ``gas'' which can be described by a polytropic equation of
state.
Previous studies showed that such an approximation adequately describes the
mechanism of the growth of global modes, but the effects of the
multi-component nature of the galactic disks, and especially the
presence of the cold gas component might be important (Lin \& Shu 1966;
Sellwood \& Carlberg 1984; Bertin \& Romeo 1988).
In this paper we study the dynamics of the disk of NGC 1566 using both
one-component, and multi-component models taking the effects of star
formation and the self-gravity of gas into account.

{\it One-Component Disk.} The behavior of our one-component model is
described by the standard set of the continuity equation, the momentum
equations, and Poisson's equation in polar coordinates:
\be
{\partial u \over \partial t}
 + u {\partial u \over \partial r}
 + {v \over r} {\partial u \over \partial \phi}
 - {v^2 \over r} =
 - {1 \over \sigma} {\partial P_s \over \partial r}
 - {\partial \over \partial r} \Big( \Psi + \Psi _H \Big)
\ ,
\label{eq10}
\ee
\be
{\partial v \over \partial t}
 + u {\partial v \over \partial r}
 + {v \over r} {\partial v \over \partial \phi}
 + {v u \over r} =
 - {1 \over \sigma r} {\partial P_s \over \partial \phi}
 - {1 \over r} {\partial \over \partial \phi} \Big( \Psi + \Psi_H \Big)
\ ,
\label{eq11}
\ee
\be
 {\partial \sigma \over \partial t}
 + {1 \over r} {\partial \over \partial r} \Big( r\sigma u \Big)
 + {1 \over r} {\partial \over \partial \phi} \Big( \sigma v \Big) = 0
\ ,
\label{eq12}
\ee
and
\be
\Psi(r,\phi) = -
\int_{R_{in}}^{R_{out}} \int_{0}^{2 \pi}
{\sigma(r^{\prime},\phi^{\prime}) r^{\prime} dr^{\prime} d \phi^{\prime}
 \over {\sqrt{r^{2}+r^{\prime 2} -
 2rr^{\prime} \cos (\phi - \phi^{\prime})}}}
\ ,
\label{eq13}
\ee
Here, $u$ and $v$ are the radial and azimuthal velocities within the
disk, and $\sigma$ is the surface density.  
The self-gravity of the gas $\Psi$, the explicit contribution $\Psi_H$ from
the rigid halo, and the pressure gradient determine the behavior of
perturbations in the disk.
All of the dependent variables are functions of the radial coordinate
$r$, the azimuthal angle $\phi$, and the time $t$.

{\it Multi-Component Disk.} To study how the cold gaseous component and
the presence of
star formation might affect the growth of spirals in the NGC 1566 disk,
we consider a multi-component model introduced by Korchagin \&
Theis (1999). 
This model splits the galactic disk into the gas, ``active'' stars, and
``inactive'' stellar remnants, considering all three components as
fluids, coupled by nonlinear interchange processes and by the common
gravitational potential.
The nonlinear interchange processes bear the basic features of the
``chemo-dynamical'' approach, developed by Theis, Burkert, \& Hensler
(1992).

We assume that the disk of NGC 1566 was gaseous in the past.
Its initial mass, surface density distribution, and rotation were taken
from today's background distributions for the stellar disk in NGC 1566.
We then assume that
the growth of spiral perturbations is accompanied by the
transformations of the gaseous disk into the ``stellar'' one 
given by the right-hand-sides of the continuity equations. 
In cylindrical coordinates they are:
\be
{D_g \sigma_g \over Dt} =
 - C_2 \sigma_g^2 + \eta {\sigma_s \over \tau}
\label{eq14}
\ee
\be
{D_s \sigma_s \over Dt} =
\zeta C_2 \sigma_g^2 - {\sigma_s \over \tau}
\label{eq15}
\ee
\be
{D_r \sigma_r \over Dt} =
(1-\zeta) C_2 \sigma_g^2 +(1-\eta) {\sigma_s \over \tau}
\label{eq16}
\ee
Here $D_{g,s,r}/Dt$ are the corresponding ``material'' time derivatives
written in the cylindrical coordinates
\be
{D_{g,s,r} \over Dt} = {\partial \over \partial t}
 + {1 \over r} {\partial \over \partial r} r u_{g,s,r}
 + {1 \over r} {\partial \over \partial \phi} v_{g,s,r}
\ ,
\label{eq17}
\ee
$\sigma_{g,s,r}$ are the surface densities, and $u_{g,s,r}$ and
$v_{g,s,r}$ are the radial and azimuthal components of the velocities of
gas, stars and remnants.

The terms on the right-hand-side of the equations (14)--(16) model the
birth of stars according to the Schmidt law ( Schmidt 1959)
with the rate proportional
to the constant $C_2$, return of a fraction $\eta$ of stellar mass into
the ISM by the ejection, and the formation of stars and remnants with
the efficiency $\zeta$ and $1-\zeta$ correspondingly.
The parameter $\tau$ represents the mean stellar lifetime of massive
stars.

With the help of the equations (14)--(17), the corresponding momentum
equations can be written as:
\be
\sigma_g {D \mbox{\boldmath $v$}_g \over Dt}
 + \nabla P_g
 + \sigma_g \nabla \Big( \Psi + \Psi_H \Big)
 = - C_2 \sigma_g^2 \mbox{\boldmath $v$}_g
 + \eta{\sigma_s \over \tau} \mbox{\boldmath $v$}_s
\label{eq18}
\ee
\be
\sigma_s {D \mbox{\boldmath $v$}_s \over Dt}
 + \nabla P_s
 + \sigma_s \nabla \Big( \Psi + \Psi_H \Big)
 = \zeta C_2 \sigma_g^2 \mbox{\boldmath $v$}_g
 - {\sigma_s \over \tau} \mbox{\boldmath $v$}_s
\label{eq19}
\ee
\be
\sigma_r {D \mbox{\boldmath $v$}_r \over Dt}
 + \nabla P_r
 + \sigma_r \nabla \Big( \Psi + \Psi_H \Big)
 = (1-\zeta) C_2 \sigma_g^2 \mbox{\boldmath $v$}_g
 +(1-\eta) {\sigma_s \over \tau} \mbox{\boldmath $v$}_s
\label{eq20}
\ee

The gravitational potential $\Phi$ is determined by the total density of 
all components, and can be written in the form of a Poisson integral as:
\be
\Psi(r,\phi) = -\int_{R_{in}}^{R_{out}} \int_{0}^{2 \pi}
{ (\sigma_g(r^{\prime},\phi^{\prime})
 + \sigma_s(r^{\prime},\phi^{\prime})
 + \sigma_r(r^{\prime},\phi^{\prime}))
r^{\prime} dr^{\prime} d\phi^{\prime} \over
{\sqrt{r^{2}+r^{\prime 2} - 2rr^{\prime} \cos (\phi-\phi^{\prime})}}}
\label{eq21}
\ee
The equations of state for the ``pressures'' $P_{g,s,r}$ of the three
components close the system of equations (14)--(21):
\be
P_{g,s,r} = K_{g,s,r} \sigma_{g,s,r}^{\gamma_{g,s,r}}
\label{eq22}
\ee

The parameters used in the numerical simulations are listed in the Table 1.

\section{COMPUTATIONAL METHODS}

\subsection{Linear Modal Analysis}

A linear modal analysis starts by considering
non-axisymmetric perturbations of the equations (10)-(13), in the form of a
Fourier decomposition:
\be
f( r ) + f_1( r ) e^{i m \phi- i \omega t} ~.
\label{eq23}
\ee
Here, $f( r )$ denotes any unperturbed quantity, $m$ is the azimuthal
wave number and $\omega$ is the complex frequency of the perturbation.
Following the commonly used procedure (e.g., Lin \& Lau 1979; Adams et
al. 1989), a governing integro-differential equation can be obtained
after the substituting expression (23) into the equations (10)--(13).
To eliminate the singularity at the the origin $r = 0$, we introduce the
coordinate transformations of the perturbed enthalpy $w_1$, potential
$\psi_1$, and the surface density $\sigma_1$ as:
$$
w_1      = r^m \tilde{w}_1     ~, \quad
\Psi_1   = r^m \tilde{\Psi}_1  ~, \quad
\sigma_1 = r^m \tilde{\sigma}_1~.
$$
After linearizing equations (10)--(13) with respect to the perturbed
quantities, and eliminating perturbed velocities, one obtains an
equation of the form:
\begin{equation}
\frac{d^2}{dr^2} (\tilde{w}_1 + \tilde{\Psi}_1)
+ A \frac{d}{dr} (\tilde{w}_1 + \tilde{\Psi}_1)
+ B (\tilde{w}_1 + \tilde{\Psi}_1)
- \frac{D}{c_r^2} \tilde{w}_1
= 0 ~.
\label{eq24}
\end{equation}
Here, the coefficients $A$, $B$, and $D$ are given by the expressions:
\be
A = \frac{2 m + 1}{r}
+ \frac{1}{\sigma}\frac{d\sigma}{dr}
- \frac{1}{D} \frac{dD}{dr}~,
\label{eq25}
\ee
\be
B = \frac{m}{r}
\left[ \left( \frac{1}{\sigma}\frac{d\sigma}{dr}
- \frac{1}{D} \frac{dD}{dr} \right)
\left( 1 - \frac{2 \Omega}{\omega- m \Omega} \right)
- \frac{2}{\omega- m \Omega}\frac{d\Omega}{dr} \right]~,
\label{eq26}
\ee
\be
D = \kappa^2 - (\omega - m \Omega)^2~.
\label{eq27}
\ee
In these equations, $\Omega$ is the angular velocity and $\kappa$ is the 
epicyclic frequency defined as
$$
\kappa^2 = \frac{1}{r^3} \frac{d}{dr} (r^4 \Omega^2)~.
$$
The perturbed potential $\tilde{\Psi}_1$ is expressed by the Poisson
equation in integral form:
\begin{eqnarray}
& &
\tilde{\Psi}_1( r )
= - 2 \pi G \frac{\Gamma(m+1/2)}{\Gamma(m+1)\Gamma(1/2)}
\nonumber \\
& & \qquad
\left[
\int_0^r \tilde{\sigma}_1( r' ) \left( \frac{r'}{r} \right)^{2m+1}
F \left( \frac{1}{2},m+\frac{1}{2};m+1;\frac{r'^2}{r^2} \right) dr'
\right.
\nonumber \\
& & \qquad
\left.
+ \int_r^{R_{out}} \tilde{\sigma}_1( r' )
F \left( \frac{1}{2},m+\frac{1}{2};m+1;\frac{r^2}{r'^2} \right) dr'
\right]~,
\label{eq28}
\end{eqnarray}
where $F$ is a hypergeometric function.
The perturbed enthalpy $w_1$ and surface density $\sigma_1$ are related
by $w_1 = ( c_r^2/\sigma ) \sigma_1$.

The inner and outer
boundary conditions, together with the equations (24) and (28)
formulate the eigenvalue problem.
The inner boundary condition follows from the regularity of solutions 
at the center of the disk:
\be
\frac{d}{dr} (\tilde{w}_1 + \tilde{\Psi}_1)
+ \frac{m}{2 m + 1}
\left[ \left( \frac{1}{\sigma} \frac{d\sigma}{dr}
- \frac{1}{D}\frac{dD}{dr} \right)
\left( 1 - \frac{2\Omega}{\omega - m \Omega} \right)
-\frac{2}{\omega - m \Omega}\frac{d\Omega}{dr}\right]
(\tilde{w}_1 + \tilde{\Psi}_1)
= 0 ~.
\label{eq29}
\ee
Likewise, at the outer boundary where $c_r^2 = 0$, we require
\be
\frac{1}{\sigma}\frac{dP_s}{dr}
\left[ \frac{d}{dr} (\tilde{w}_1 + \tilde{\Psi}_1)
+ \frac{m}{r} (\tilde{w}_1 + \tilde{\Psi}_1) \right]
- D \tilde{w}_1 = 0 ~.
\label{eq30}
\ee
The eigenvalue problem formulated by the equations (24)--(28), and the
boundary conditions (29) and (30) is solved numerically by means of a
matrix method.
In this method, the perturbed enthalpy $\tilde{w}_1$ is represented by
the $(N+1)$-dimensional vector with its value taken at $(N + 1)$ radial
grid points.
When 
the derivatives in equation (24) are expressed as finite differences, and
the integral in the equation (28) is expressed by a finite sum, the problem is
reduced to an $(N + 1) \times (N + 1)$ matrix equation:
\be
\sum_{k=0}^{N} M_{ik} \tilde{w}_{1,k} = 0~,~{\rm for}~i=0, 1, ..., N
\label{eq31}
\ee
In practice, the first and last rows in the equation (\ref{eq31}) are
replaced by the boundary conditions expressed in proper differenced form.

Equation (31) has a non-trivial vector solution if the condition
\be
{\rm det}M(\omega) = 0
\label{eq32}
\ee
is satisfied, which yields an eigenfrequency $\omega$ of the global
mode.
The growth rate and the pattern speed of the mode are then written as
${\rm Im}(\omega)$ and by ${\rm Re}(\omega)/m$ respectively.

\subsection{2D Numerical Simulations}

To solve the one-component, and multi-component hydrodynamical equations we
use two-dimensional numerical codes based on a second order Van Leer
advection scheme implemented by Stone \& Norman (1992) in a general
purpose fluid dynamics code, called ZEUS-2D.
This code was designed for modeling astrophysical systems in two spatial
dimensions, and it can be used for the simulations in a variety of
astrophysical processes.
The ZEUS-2D code uses accurate enough hydrodynamical algorithms and
allows to include complex physical effects in a self-consistent 
fashion. This code thus provides a good basis for the implementation of the
nonlinear mass transfer processes into the multi--phase hydrodynamics.

In our code, the mass and momentum interchange processes between disk
components are computed at the first sub-step of the ZEUS-type code.
To advance the solutions due to interchange processes given by the
right-hand sides of the equations (14)--(16) we use a fifth order
Cash-Karp Runge-Kutta routine with the time step limitation imposed by
the Courant-Friedrichs-Levy criterion and the values of the parameters
$\tau$ and $C_2$ in the mass and momentum interchange processes.

In both codes, the Poisson equation is solved by applying the
two-dimensional Fourier convolution theorem in polar coordinates
(Binney \& Tremaine 1987).
By introducing the new variable $u = \ln r$, the Poisson integral in
equation (21) can be rewritten as:
\begin{equation}
\Psi'( u,\phi ) =
- \int_{\ln R_{in}}^{\ln R_{out}} \int_0^{2\pi}
\sigma'( u', \phi' ) K( u - u', \phi - \phi' ) du' d\phi'~,
\label{eq33}
\end{equation}
where $\Psi' = \Psi e^{u/2}$ is the reduced potential,
$\sigma' = \sigma e^{3u/2}$ is the reduced surface density,
and the Poisson kernel is defined by the expression
$K( u,\phi ) = 1/\sqrt{2(\cosh u - \cos \phi)}$.
The Poisson kernel in equation (33) depends on the differences
$u-u'$ and $\phi-\phi'$, and therefore the Fourier convolution theorem
can be applied to obtain the reduced potential $\Psi'$.

The codes solve hydrodynamical equations using equally spaced azimuthal,
and logarithmically spaced radial zones. For the  one-component
simulations a grid with $512\times512$ zones was employed.
The simulations of the dynamics of multi-component disk were done with a
grid of $256\times256$ zones.
All simulations were performed on the parallel supercomputer VPP300/16R
at the National Astronomical Observatory of Japan.

\section{LINEAR MODES }

Figure 5 shows the equilibrium curves of our ``standard'' model for 
NGC 1566 disk.
These curves were computed for the surface density distribution,
velocity dispersion, and the rotation curve in the disk given by the
expressions (2), (3) and (4), and for the ratio of the $z$-velocity
dispersion to the radial one equal to 0.6.
The parameters of the rotation curve were chosen to imitate the most
probable ``medium'' rotation curve of NGC 1566.

With these parameters, the minimum value of the Toomre $Q$-parameter
defined by the equation (5) is equal to 2.45.
The procedure of searching of the linear global modes, described in the
section 5.1, did not reveal any unstable modes.
This result is in apparent contradiction with the existence of the
spirals in NGC 1566, and we have searched therefore for a possible
unstable configuration within observational errors.
The unstable equilibrium configuration can be achieved if the disk has
a slower rotation and/or a smaller radial velocity dispersion.
We calculated the equilibrium properties of the disk with the ``low''
rotation curve with other equilibrium parameters being unaltered. 
In this disk, the $Q$-parameter has minimum value about 1.65. Nevertheless,
the linear modal analysis still did not reveal any unstable global modes.
This result concurs with the stability analysis by Vauterin \& Dejonghe
(1996) of self-gravitating disks, who found that a high enough value of
the central velocity dispersion stabilizes the disk.
Comparison of their Figure 1 and Figure 12 shows that the disk is stable
when the ratio of the central velocity dispersion to the value of the
rotational velocity in its ``flat'' part is $\lesssim 0.8$.
With the ``low'' rotation curve, the ratio of the central velocity
dispersion to the maximum value of the rotational velocity is about 1.0,
which stabilizes the system.

Another way to achieve an unstable configuration is to increase the ratio
$c_z/c_r$, and hence decrease the radial velocity dispersion.
N-body experiments by Lacey (1984) predict the value of the ratio of the
velocity dispersions  $c_z/c_r$ to be about 0.8 which is larger than
the value $c_z/c_r \simeq 0.6$ found by Villumsen (1985).
We studied the stability properties of the disk which has a ``medium''
rotation curve and a ratio of the velocity dispersions $c_z/c_r=0.8$.
The minimum value of $Q$-parameter in such a model is lower than that in
the ``standard'' disk ($\sim 1.8$),  but the linear stability analysis
shows that such a disk is stable.

The situation changes, however, if we choose the ``low'' rotation curve.
The matrix method described in the previous section yields an
$m=2$ global mode with eigenvalues 
${\rm Re}( \omega_2 ) = 0.893$, ${\rm Im} (\omega_2) = 0.347$
as the most unstable global mode.
Its nearest competitor, a three-armed spiral, has eigenvalues equal to
${\rm Re} (\omega_3) = 1.426$, and ${\rm Im} (\omega_3) = 0.156$.

Similar behavior occurs in a disk with a velocity dispersion ratio
increased to $c_z/c_r = 1.0$.
We found that this disk is unstable with the most probable ``medium''
rotation curve.
Figure 6 presents the equilibrium properties of such a disk, and Figures 7
and 8 show contour plots for dominant $m=2$, and its nearest competitor, $m=3$ global
modes. The contour levels in Figures 7 and 8 are logarithmically spaced
between the maximum value of the perturbed density, and one-hundredth
of the maximum perturbed density.
The eigenvalues of these two modes found by the matrix linear method are
${\rm Re} (\omega_2) = 1.567$, ${\rm Im} (\omega_2) = 0.280$,
and ${\rm Re} (\omega_3) = 2.859$, ${\rm Im} (\omega_3) = 0.217$, respectively.

In summary, the results of our linear modal analysis allow us to conclude
that the disk of NGC 1566 is unstable if the ratio of the vertical and the
azimuthal velocity dispersions is close to unity.
With the ratio $c_z/c_r$ equal to unity, the disk is unstable with the 
observationally most probably ``medium'' rotation curve.
If the disk of NGC 1566 has a lower value of this ratio, the actual
rotation curve of the disk of NGC 1566 must be close to the lower limit
determined by the observational errors.
The $m=2$ spiral, and the $m=3$ spiral are the most unstable in the disk,
and the dominating $m=2$ global mode resembles open
spiral arms in the inner regions of NGC 1566. 

Linear modal analysis is unable, however, to determine the relative
amplitudes of the two competing unstable modes at their nonlinear saturated 
stage.
Furthermore, 
it can not predict the radial dependence of the amplitude
actually observed in spirals.
These questions are addressed by the nonlinear two-dimensional
simulations which we discuss in the following sections.

\section{NONLINEAR SIMULATIONS}
 
\subsection{Nonlinear Dynamics of a ``Standard'' Disk}

The results of the linear modal analysis are well complemented by direct
two-dimensional simulations of disk dynamics.
The time dependence of the global modes in the disk
can be expressed in terms of the global Fourier
perturbation amplitudes, which are defined as:
\be
A_m \equiv {1 \over {M_{d}}}
\left\vert \int_{0}^{2 \pi} \int_{R_{in}}^{R_{out}}
\sigma(r,\phi) r dr \, e^{-im\phi} d\phi \right\vert
\, .
\label{eq34}
\ee
Here $M_d$ is the mass of the disk.

Figure 9 plots the time dependence of the global Fourier amplitudes for
modes $m = 1 - 6$ resulting from a random initial density
perturbation seeded into every grid cell of the 
``standard'' disk with a ratio of dispersions $c_z/c_r = 0.6$. 
Within each cell, the perturbation can have a value of up to
one part in a thousand of the 
equilibrium density.
Figure 9 confirms the results of the linear modal analysis.
The ``standard'' disk with a ratio of dispersions $c_z/c_r = 0.6$ is
stable for the ``medium'' rotation curve.
Similar stable behavior of the global modes is observed in the disk with
``low'' rotation.
Note that random perturbations at the level $\sim 10^{-6}$ exist in the
disk during the entire computational timeframe. 
This low level of departure from the equilibrium state does not lead to the
growth of the swing-amplified spirals.

Our previous consideration of the stability properties of galaxy NGC 1566
was based on a one-component approximation.
An additional 
cold gas component may play an important role in the destabilization of
self-gravitating disks.
In linear approximation, this effect has been studied by number of authors,
beginning with the work of Lin \& Shu (1966) and Lynden-Bell (1967) (see
also Jog \& Solomon 1984; Sellwood \& Carlberg 1984;
Bertin \& Romeo 1988).
Recently, Korchagin \& Theis (1999) extended the realization of the
destabilizing role of a cold gas component onto multi-phase
star-forming disks. They demonstrated that spiral structure will grow
faster on a non-stationary star-forming environment in comparison to a
one-component system with the same total surface density distribution
and rotation curve.
It might be possible therefore, that the presence of gas will change the
stability properties of a ``standard'' disk.

To study this possibility, we simulated the dynamics of the
multi-component model described in the Section 4.
We started our simulations by assuming that the multicomponent disk is
initially in a gaseous phase with a small admixture of stars, 
and that the gaseous disk has a surface density distribution which 
corresponds to the observed density profile in NGC 1566. 
The rotation curve of a gaseous disk was assumed to be the most probable
``medium'' curve for the galaxy NGC 1566.
Figure 10 shows the equilibrium properties of the gaseous component
computed at the initial moment of time. The minimum value of the
$Q$-parameter for the gaseous phase $Q = 1.22$, and
hence, the multi-component disk is unstable to spiral
perturbations.

This disk was also seeded with random initial density perturbations of
amplitude up to one-thousandth of the equilibrium density.
The subsequent behavior of this disk is determined by the mass and
momentum transformations given by the right-hand-sides of equations
(14)--(16), and by the development of the random initial perturbations.
The two-dimensional simulations demonstrate that the
multi-component disk is indeed unstable, which is illustrated by the global
amplitude diagram (Figure 11) plotted for the density perturbations
growing in the stellar component.
The growth of the Fourier harmonics is accompanied by the transformation
of the gas phase into the stellar one as shown in Figure 12.
The main unstable mode has an azimuthal wavenumber $m = 2$, but the growth
of perturbations is stabilized at a rather low level
$\log_{10} A_m \sim -2$.
More importantly, the growing perturbations do not resemble the spiral arms
observed in the disk of NGC 1566.
Figure 13 displays contour plots of the density perturbations in the
gaseous phase taken at a time $t = 20$.
In the central region we observe a short open spiral outlined by the shock
fronts, which is embedded in a system of weak ring-like structures.
At later stages of evolution, when most of the gas is transformed into
the stellar component, the density perturbation in the stellar component
resembles a short bar-like perturbation rather then a global spiral
structure (Figure 14).

\subsection{``Cold'' Disks}

The results of the linear modal analysis described in Section 6 show
that NGC 1566 is unstable with respect to the growth of a two armed spiral
if the
ratio of the velocity dispersions, $ c_z/c_r$, is close to unity.
Figure 15 plots the behavior of the global amplitudes for the Fourier
harmonics $m = 1 - 6$ developing from random perturbations in the disk
with the ratio of the velocity dispersions $c_z/c_r$ equal to unity.
In agreement with the linear modal analysis, the nonlinear simulations show
that the $m=2$ global mode is the primary instability of the disk.
The dominant two-armed mode experiences exponential growth until time $t
= 40$, when the exponential growth phase merges into a lingering
saturation phase, at which point the $m=2$ mode saturates at an amplitude
level of $\log_{10} A_2 \approx -0.5$.

The modal growth rates for the $m=2$ perturbation (as well as its nearest
competitor, the $m=3$ mode) observed in the nonlinear simulations are equal to
${\rm Im} (\omega_2) = 0.27$ and ${\rm Im} (\omega_3) = 0.21$, which is
in a good agreement with the results of the linear modal analysis
${\rm Im} (\omega_2) = 0.280$, ${\rm Im} (\omega_3) = 0.217$.
Figure 16 shows a contour plot of the surface density perturbation
taken at time $t=30$ while the perturbations were experiencing the linear
phase of exponential growth. Similarly to Figures 7 and 8, contours
are logarithmically spaced between the maximum value of perturbed
density and one-hundredth of the maximum density perturbations.
Figure 16 further demonstrates that the two-armed open spiral, which
looks similar to the one obtained in the linear global analysis (see
Figure 7), emerges from the random perturbations seeded in the unstable
disk with $c_z/c_r = 1$.

During the saturation phase, when the exponential growth of the spiral
pattern is stopped, the spiral pattern deforms from its originally
sinusoidal shape.
Figure 17 shows contour plots for the density perturbations in the
disk taken at time $t=45$.
These contour plots depict the formation of the shock front at the concave
edge of the spiral arms.
However, the perturbation maintains itself as a two-armed spiral during the
entire simulation duration.

In Section 2, we discussed the observed surface brightness variations in
the spiral arms of galaxy NGC 1566.
To compare the amplitude variations of the perturbations seen in
the numerical simulations with the observed surface brightness
variations of the NGC 1566 spirals, we plotted in Figure 18 the azimuthal
variations of the function defined as $ -2.5 \log_{10} \sigma(t,R_N,\phi)$,
which was taken at time $t=45$ at the fixed radii $R_N$ equal to 0.25,
0.5, 1.0, 1.5 and 2.5.
Figure 18 should be compared directly with Figure 3.
The overall behavior seen in both Figures is quite similar.
Near the center of the disk, the spiral perturbation has small amplitude 
variations.
The ``surface brightness'' variations shown in Figure 18 increase with
radius out to approximately $r=2$, but at larger radii $ r \gtrsim
2.0$, the ``surface brightness'' variations tend to decrease.
The surface brightness variations in the spiral arms of NGC 1566 do not
show a tendency to decrease within a 8.5 kpc radius.
This fact reflects a basic discrepancy between the theoretical
predictions and the observations.
The spiral arms found in linear, and nonlinear simulations are
considerably shorter in comparison to the observed spirals in NGC 1566.

The growth of unstable global modes is accompanied by momentum and
surface density redistribution of the background axisymmetric
disk structure.
Nonlinear self-interaction of global modes (Laughlin et al. 1998), and
spiral shocks funnel matter towards the center of the disk, resulting in
a steepening of the surface density profile.
Figure 19 illustrates how the long-term evolution of spiral
perturbations affects the azimuthally averaged surface density
distribution within the disk, making it steeper than the observed
density distribution.   

The detailed kinematic study of NGC 1566 undertaken by Pence et
al. (1990) found considerable discrepancy between the observed spiral
velocity field and the theoretical predictions.
In constructing a theoretical velocity field, Pence et al. (1990) used
results of calculations by Roberts \& Hausman (1984), who studied
the evolution of colliding gas clouds moving in a fixed sinusoidal
spiral potential.
Pence et al. (1990) found that the observed velocity residuals across
the spiral arm reach a maximum at the edge of the arm, contrary to the
predictions following from the calculations of Roberts \& Hausman (1994).
The observed gradient of the velocity field was found to be approximately five
times larger than predicted.
Figure 20 plots the azimuthal profile of the residual velocity obtained
in our numerical simulations superimposed on the surface density
profile sampled at radius 1.0 at time $t=45$.
The results of the numerical simulations show qualitative agreement with
observations.
The velocity field has a strong velocity gradient associated with the
shock front.
The maximum of the residual velocity is located near the minimum of the
density distribution, but the velocity shift found in the numerical
simulations is considerably larger than the values reported by Pence et
al. (1990).


\section{CONCLUSIONS}

This paper has focused on the observational study of the spiral arms in
galaxy NGC 1566, and on the theoretical study of the stability
properties of this galaxy. 
We have applied a global modal approach and have used two-dimensional 
one-component
and multi-component simulations to study the dynamics of the self-gravitating 
disk in the galaxy NGC 1566 and we have extended our analysis to the nonlinear
stage using 2D numerical simulations. In our theoretical analysis
of the spiral structure in NGC 1566 we have not made any additional assumptions
used in previous comparisons, and we have followed the development of the spiral
arms emerging from random perturbations all the way through to the nonlinear
saturation stage. 
The general conclusions which can be made
from our work are as follows: As C.C. Lin and his collaborators
envisioned many years ago, the most unstable linear global mode
emerging from stochastic noise in a
galactic disk determines the appearance of the spiral structure
on linear, and on nonlinear stages of the evolution of the spiral
pattern. We have found a good agreement between the linear global
modal analysis of the stability of the disk of galaxy NGC 1566,
and direct 2D numerical simulations. The theoretical spiral
pattern obtained from the linear global modal analysis, and the pattern
emerging from the random perturbations in the nonlinear
simulations are in agreement, at least qualitatively, with
the observed two-armed spiral structure in the disk of NGC 1566.

We did not aim to discriminate between the two principal physical 
mechanisms explaining the physics of spiral instabilities in gravitating disks. 
We concur however with the conclusion of Shu et al (1999) that
``...there is a coexistence of the two principal mechanisms
that produce the beautiful structures that astronomers observe
in the universe of spiral galaxies''.

The specific results of our analysis are specified below. 

1. The CCD images of NGC 1566 in $B$ and $I$-bands obtained with the
Australian National University 30in telescope were used for
measurements of the radial dependence of the amplitude variations in the spiral
arms of NGC 1566.
The azimuthal variations of the surface brightness in the $I$-band increase
with radius up to $57$ \% at $100^{\prime \prime}$.
Our results concur with the previous measurements of the amplitude
variations within the spiral arms of NGC 1566.

2. The linear stability analysis of the disk of NGC 1566 made under the
standard assumption used in the galactic dynamics, namely that the
ratio of the vertical and the radial velocity dispersions is equal to
0.6, shows that disk is stable towards spiral perturbations within
observational error bars.
We confirm this conclusion with help of 2D simulations 
of the disk evolution seeded with random perturbations.

3. By increasing the $c_z/c_r$ ratio up to $0.8 - 1.0$, the disk, when
seeded with random perturbations, becomes
unstable with respect to $m=2$, and $m=3$ spiral modes.
The growth rates and the shapes of the global modes found in the linear
analysis, and those seen in the nonlinear simulations are in good
agreement.

The two-armed spiral mode prevails over its competitors, and thus
determines the behavior of perturbations during the linear, and during the
nonlinear phases (see Figures 16 and 17).
At the nonlinear stage, the two-armed spiral saturates at an amplitude
$\log_{10} A_2 \approx -0.5$.
The surface density, and the velocity residual variations in the arms
are in a qualitative agreement with observations.

4. The nonlinear phase of instability is characterized by the transport
of angular momentum towards the disk center making the surface
density distribution more steep than the observed surface brightness
profile in the disk of NGC 1566.
The spiral arms found in the linear modal analysis, and seen in the
nonlinear simulations, are considerably shorter than the observed
spiral arms in the disk of NGC 1566.
We argue therefore, that the surface density distribution in the disk of
the galaxy NGC 1566 was different in the past when spiral structure in
NGC 1566 was growing linearly.

\begin{acknowledgments}
We thank the RAPT Group of amateur astronomers (E. Pozza, A. Brakel,
B. Crooke, S. McKeown, G. Wyper, K. Ward, D. Baines, P. Purcell,
T. Leach, J. Howard, D. McDowell, A. Salmon, A. Gurtierrz) for providing
the images from the 30in telescope at the ANU's Mt. Stromlo Observatory.
VK acknowledges Prof. S. Miyama for hospitality, and National
Astronomical Observatory of Japan for providing COE fellowship.
BAP acknowledges the hospitality of the National Astronomical
Observatory of Japan. 
The computations were performed on the Fujitsu VPP300/16R at the
Astronomical Data Analysis Center of the National Astronomical
Observatory, Japan.
\end{acknowledgments}

\newpage
%


\clearpage
\begin{figure}
\plotone{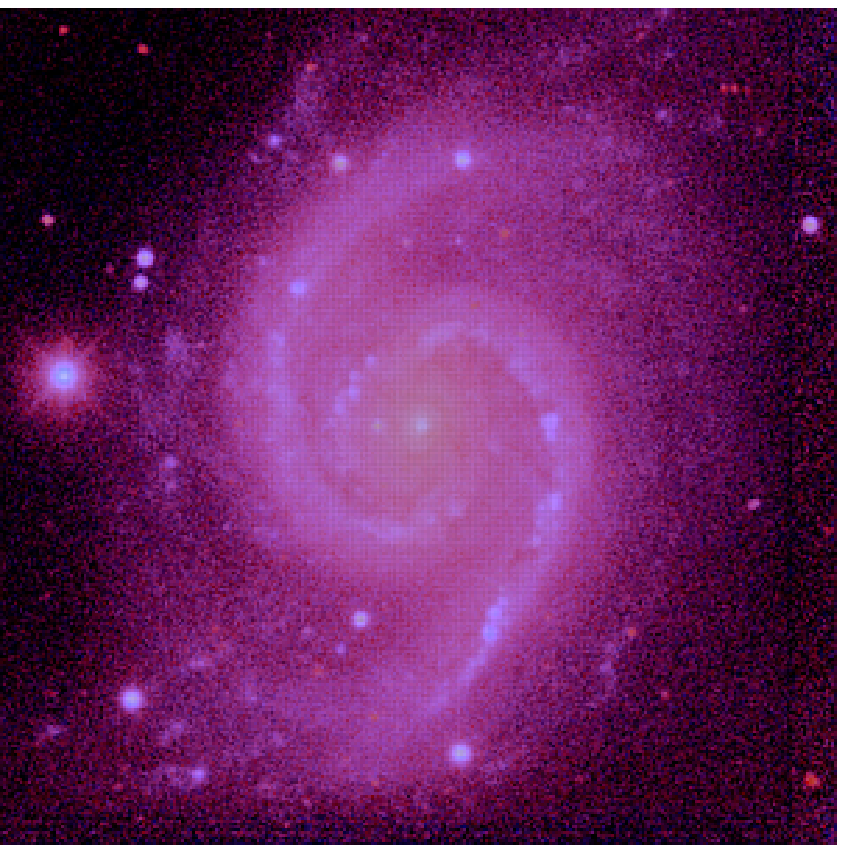}
\caption{The $B$-band image of NGC 1566 (blue)
superimposed onto the $I$-band image (red). 
South is at the top and east is to the left.
The blue HII regions are arrayed along the inside edge of the broad
arms of older stars. }
\end{figure}

\clearpage
\begin{figure}
\plotone{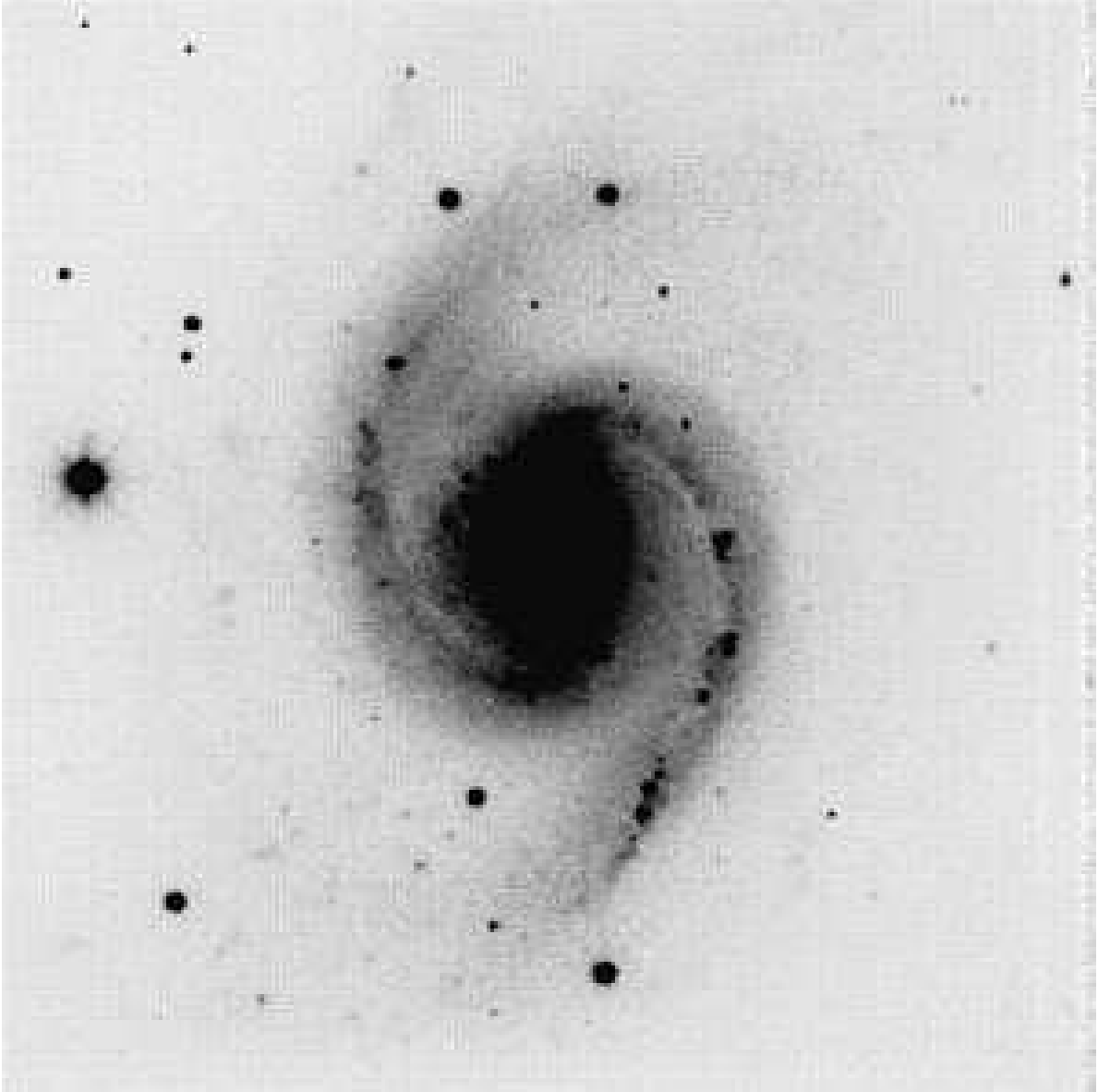}
\caption{The I-band image of NGC 1566 scanned in the
area $288^{\prime \prime}\times 288^{\prime \prime}$ with south up
and north to the left.}
\end{figure}

\clearpage
\begin{figure}
\plotone{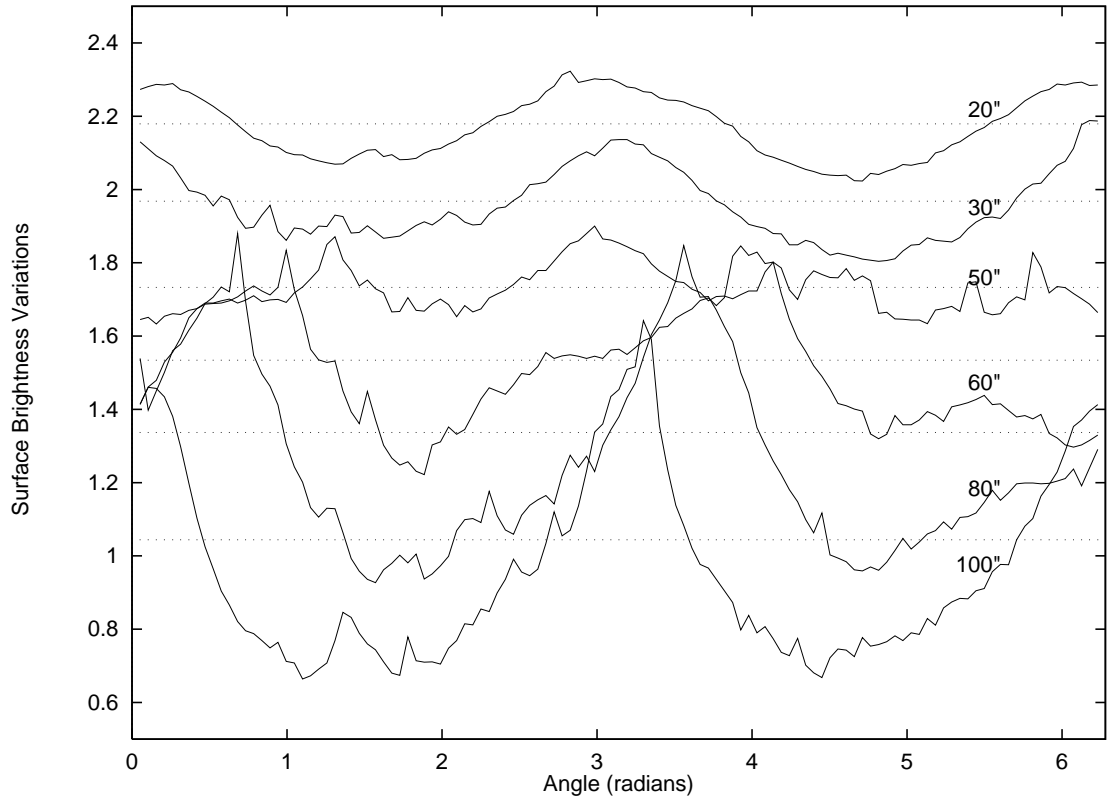}
\caption{The azimuthal variations of $\log_{10}$ 
of the number of counts 
proportional to the surface brightness plotted in the six rings with radii
$20^{\prime \prime}$, $30^{\prime \prime}$, $50^{\prime \prime}$, 
$60^{\prime \prime}$, $80^{\prime \prime}$ and $100^{\prime \prime}$.
Each ring has thickness $5.4^{\prime \prime}$.
The averaged surface brightness of each ring is shown by dotted lines.
The relative variations of the surface brightness increase with radius.}
\end{figure}

\clearpage
\begin{figure}
\plotone{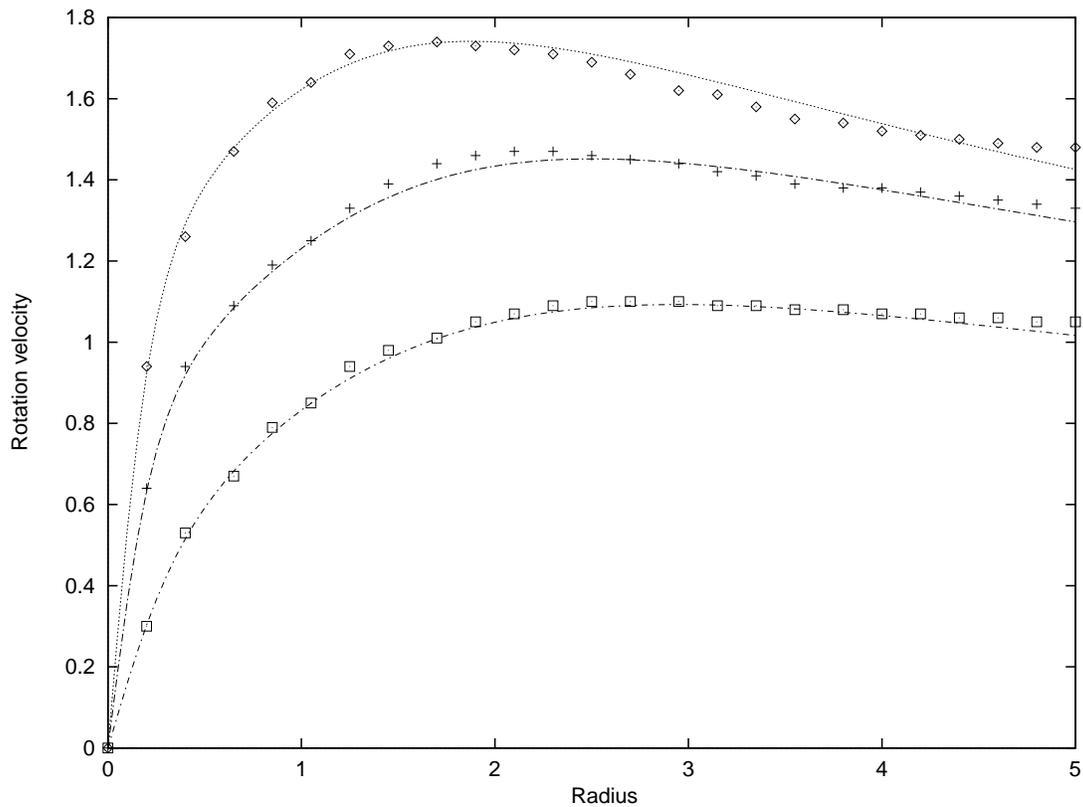}
\caption{The model of the rotation curve of NGC 1566
superimposed onto observational data taken from RB.
The ``upper'' curve (diamonds), and the ``lower'' curve (squares) arise
from the $\pm 5^{\circ}$ uncertainty in the determining of inclination
of NGC 1566. Radius and velocity are in the dimensionless units
determined in Section 3 ($V_{unit} = 149$ km/s , $L_{unit}=2$ kpc).}
\end{figure}

\clearpage
\begin{figure}
\plotone{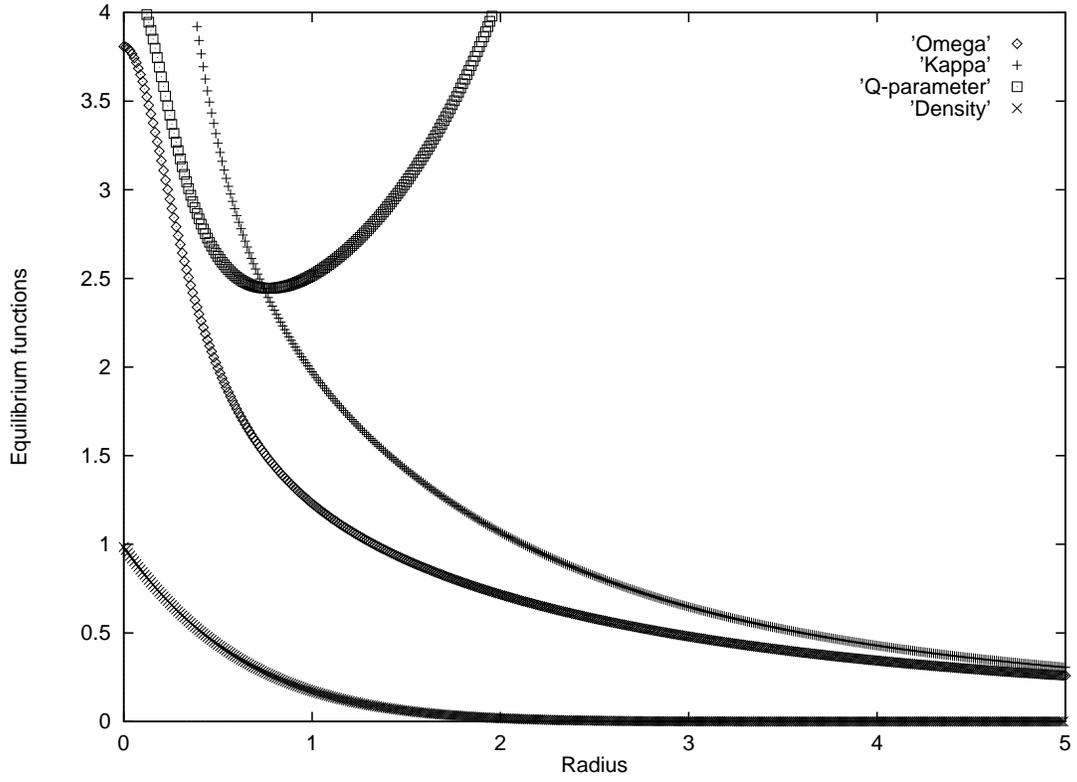}
\caption{Equilibrium curves of the ``standard'' disk model
of NGC 1566 which has the ``medium'' rotation curve, and the ratio of the
velocity dispersions $c_z/c_r = 0.6$. The disk is stable with respect to spiral
perturbations. All values are given in units $G=1, M_{unit}=10^{10} M_{\odot}$
and $L_{unit}=2$ kpc.}
\end{figure}

\clearpage
\begin{figure}
\plotone{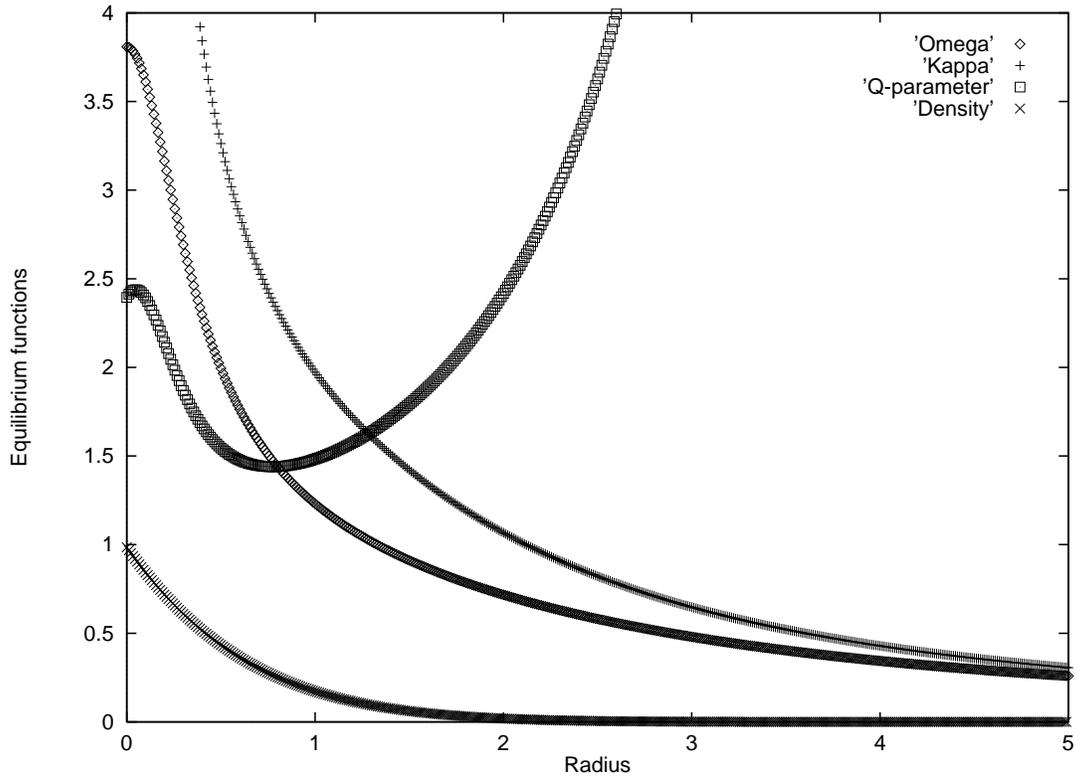}
\caption{Equilibrium curves of the unstable model with the
ratio of the velocity dispersions $c_z/c_r = 1.0$, and with the most
probable ``medium'' rotation curve.
The disk is unstable to the growth of an $m=2$ global mode. Units are the same
as in Figure 5.}
\end{figure}

\clearpage
\begin{figure}
\plotone{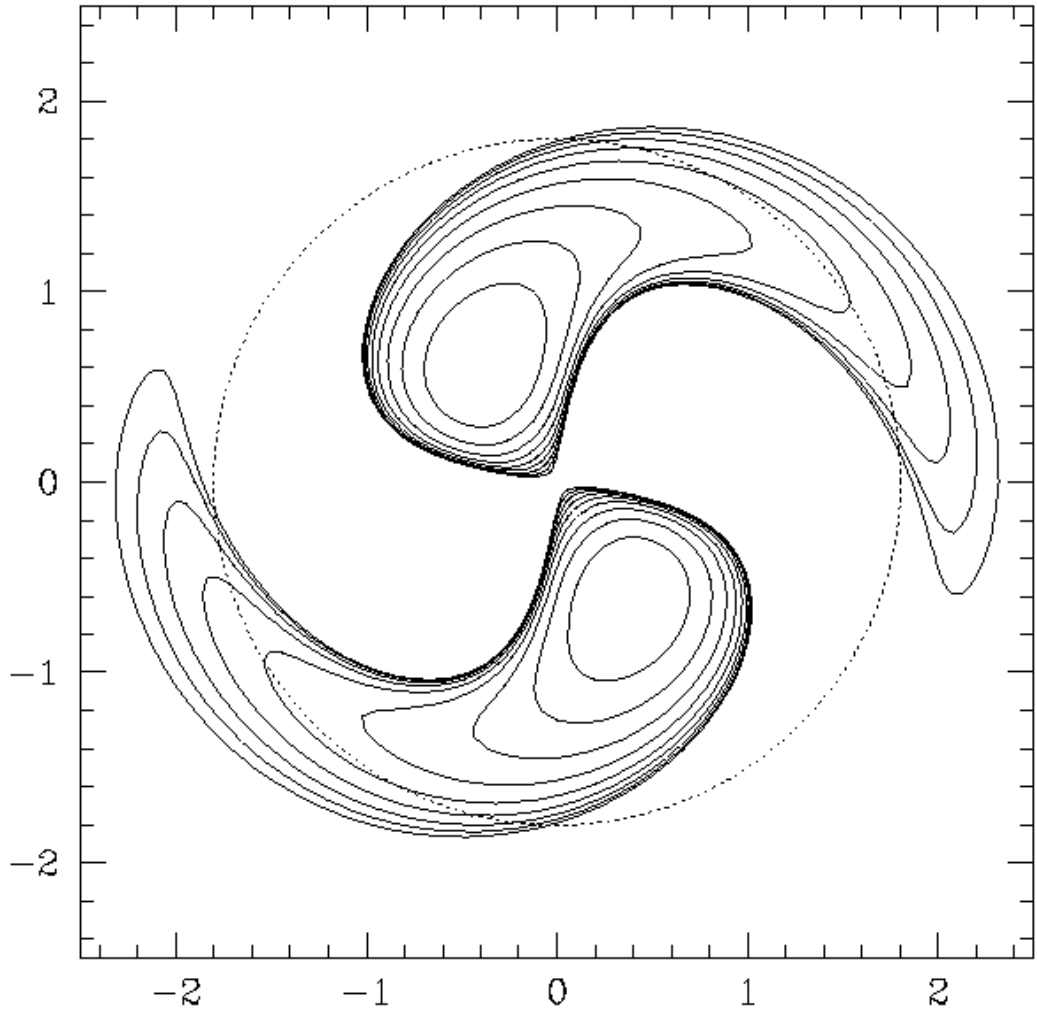}
\caption{The dominant $m=2$ global mode found in the linear 
stability analysis of the unstable disk with the equilibrium properties
shown in Figure 6. 
The eigenfrequency of the mode is ${\rm Re} (\omega_2) = 1.567$,
${\rm Im} (\omega_2) = 0.280$. $T_{unit}=1.34 \times 10^7$ years. 
Distance is measured in units $L_{unit}=2$ kpc.}
\end{figure}

\clearpage
\begin{figure}
\plotone{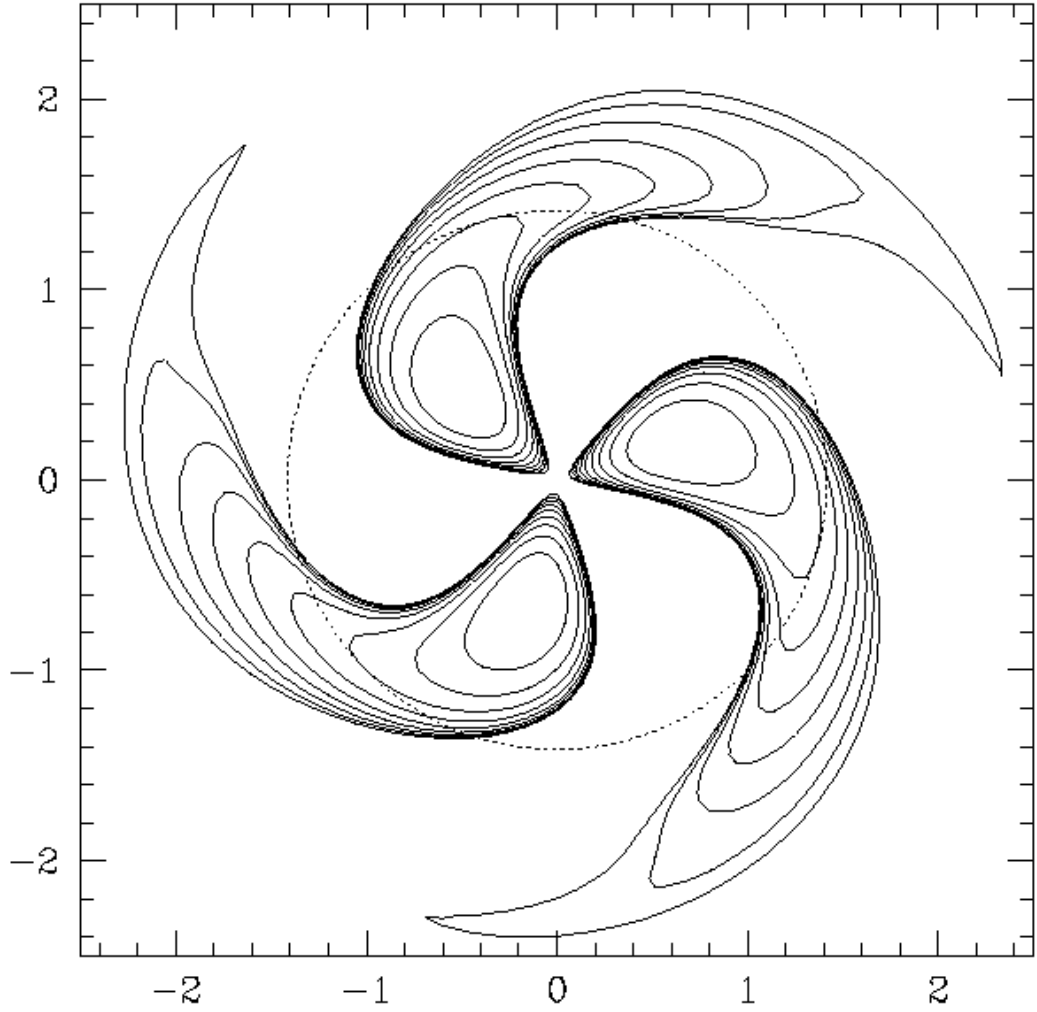}
\caption{The contour plot of the $m=3$ unstable mode in the
disk with equilibrium shown in Figure 6.
${\rm Re} (\omega_3) = 2.859$, ${\rm Im} (\omega_3) = 0.217$.
$T_{unit}=1.34 \times 10^7$ years.
Distance is in units $L_{unit}=2$ kpc.}
\end{figure}

\clearpage
\begin{figure}
\plotone{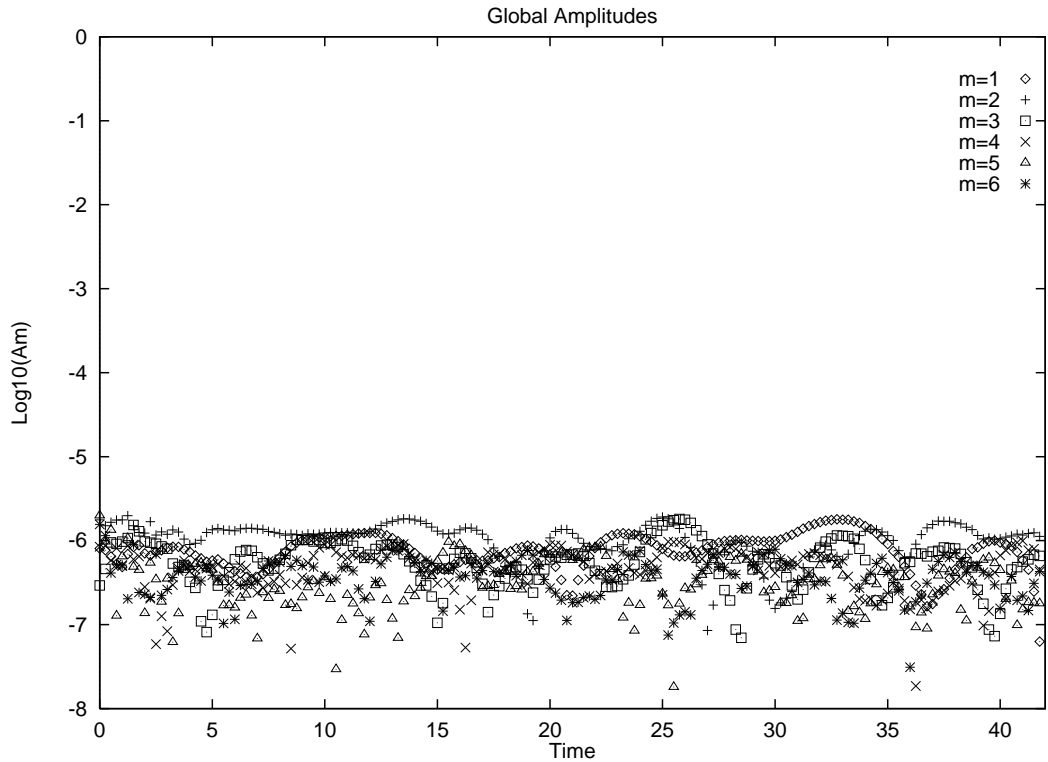}
\caption{Time dependence of the $\log_{10}$ of
the dimensionless global Fourier amplitudes determined by the equation (34)
for the spiral modes $m= 1 - 6$ computed in the ``standard'' disk with
equilibrium properties shown in Figure 5. Time is measured in units
$1.34 \times 10^7$ yrs.
The disk is stable in accordance with the linear modal analysis.}
\end{figure}

\clearpage
\begin{figure}
\plotone{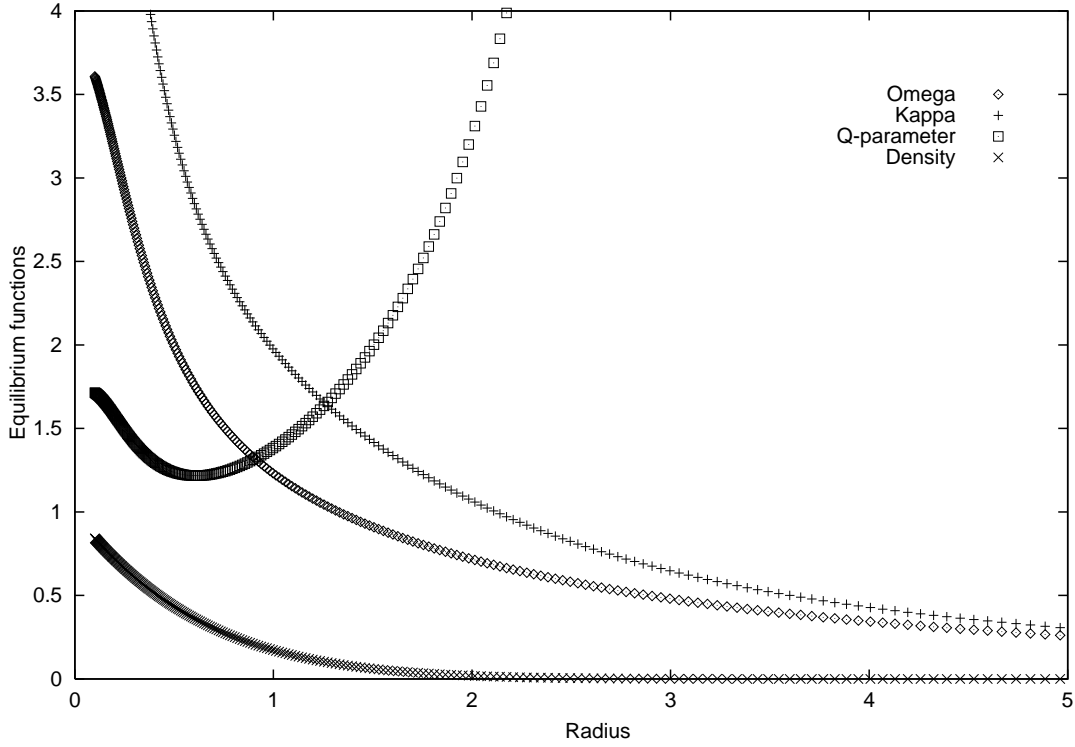}
\caption{The equilibrium curves of the gaseous component of
a multi-component disk computed at the initial moment of time.
The initial mass of the disk is equal to the mass of stellar disk in NGC
1566, and the disk has ``medium'' rotation curve.
All values are in units $G=1, M_{unit}=10^{10} M_{\odot}, _{unit}=2$ kpc.}
\end{figure}

\clearpage
\begin{figure}
\plotone{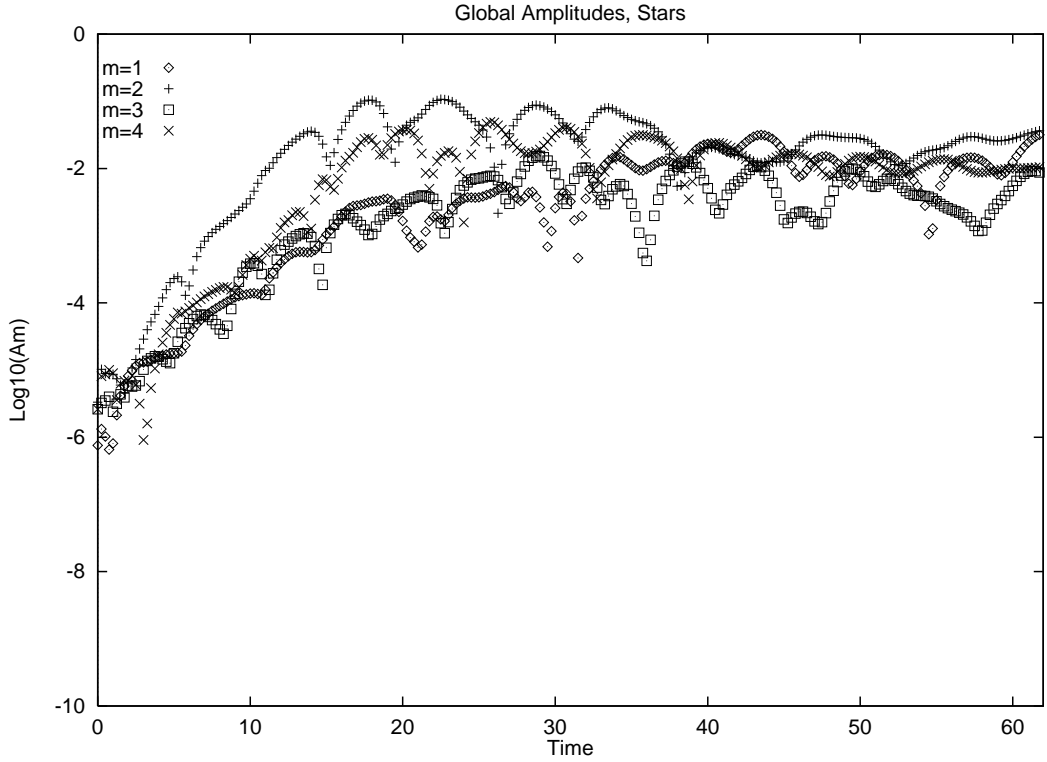}
\caption{The time evolution of the $\log_{10} A_m$ for
$m = 1 - 4$ global Fourier amplitudes determined by the equation (34).
Perturbations develop in a stellar
component of a multicomponent disk seeded with random
perturbations. The most unstable mode is the $m=2$ spiral. The
instability saturates at a low level $\log_{10} A_m \sim -2$ .
Time is in units of $T_{unit}=1.34 \times 10^7$ years.}
\end{figure}

\clearpage
\begin{figure}
\plotone{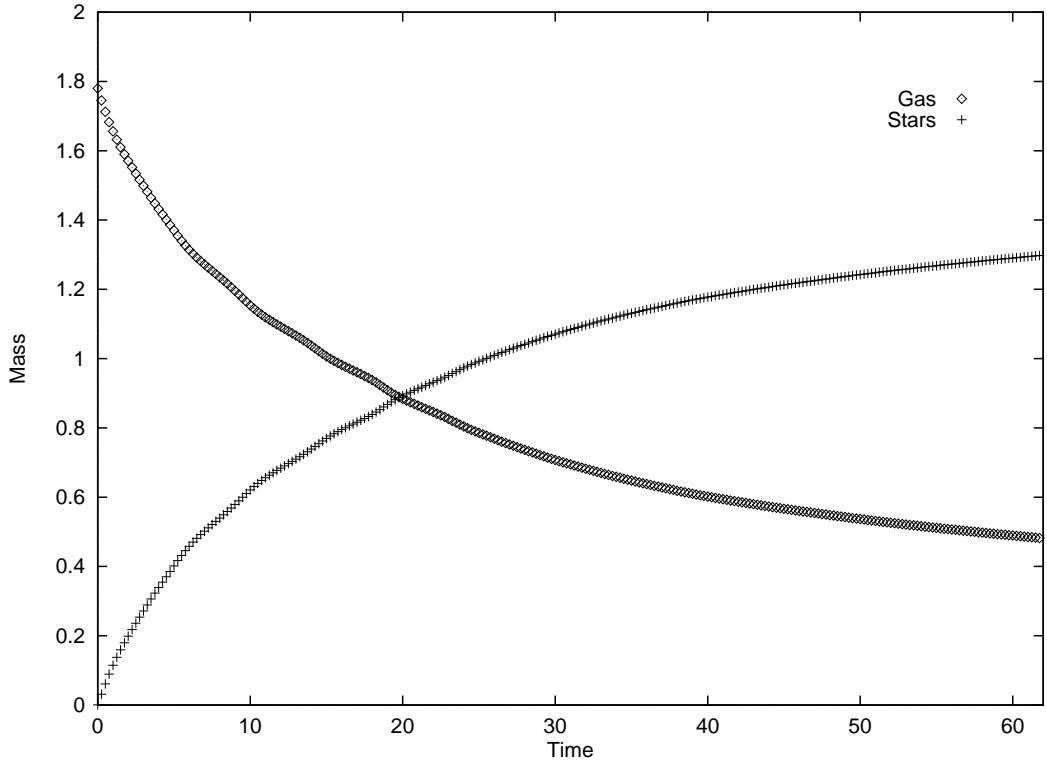}
\caption{Mass transformations between the gaseous and the
stellar components in a multi-component disk.
By the end of the numerical simulation, most of the gas has been transformed
into stars. Mass is in units of $M_{unit}=10^{10} M_{\odot}$, and
time is in units of $1.34 \times 10^7$ years.}
\end{figure}

\clearpage
\begin{figure}
\plotone{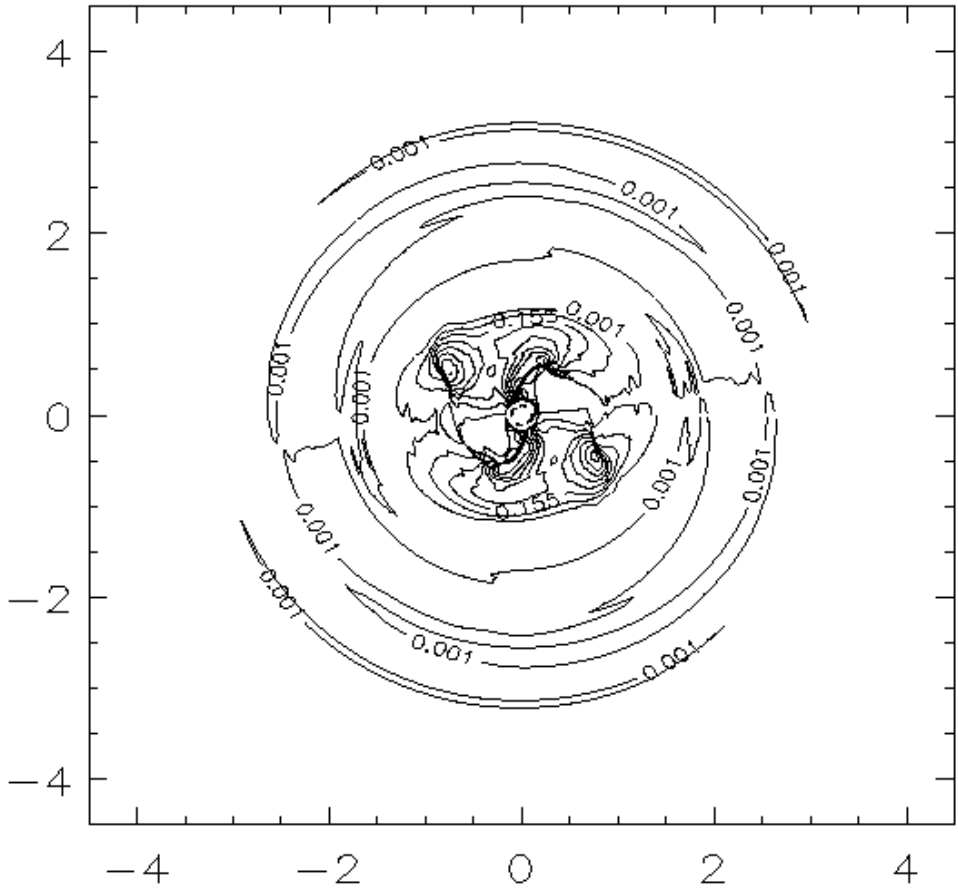}
\caption{Snapshot of the density distribution in the gaseous
phase of a multicomponent disk taken at time $t=20$.
$T_{unit}=1.34 \times 10^7$ years.
Distance is in units $L_{unit}=2$ kpc.}
\end{figure}

\clearpage
\begin{figure}
\plotone{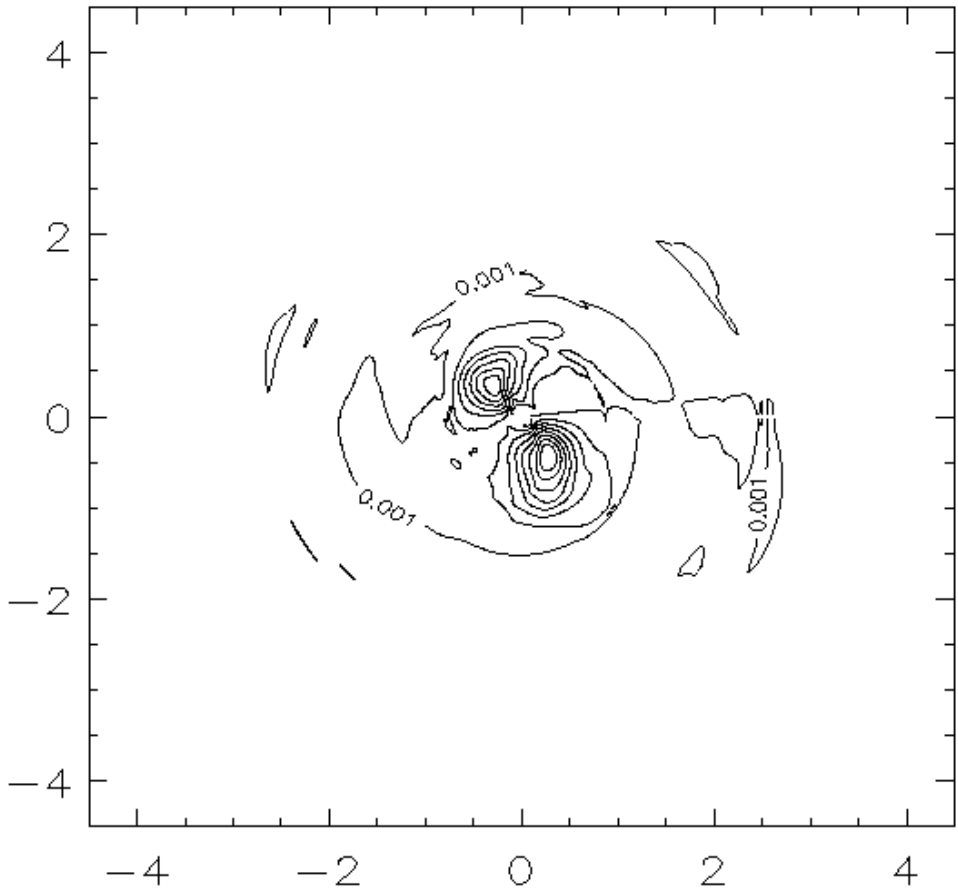}
\caption{Snapshot of the density distribution in the stellar
component of the multicomponent disk taken at time $t=40$ when most of gas
has been transformed into stars.
$T_{unit}=1.34 \times 10^7$ years.
Distance is in units $L_{unit}=2$ kpc.} 
\end{figure}

\clearpage
\begin{figure}
\plotone{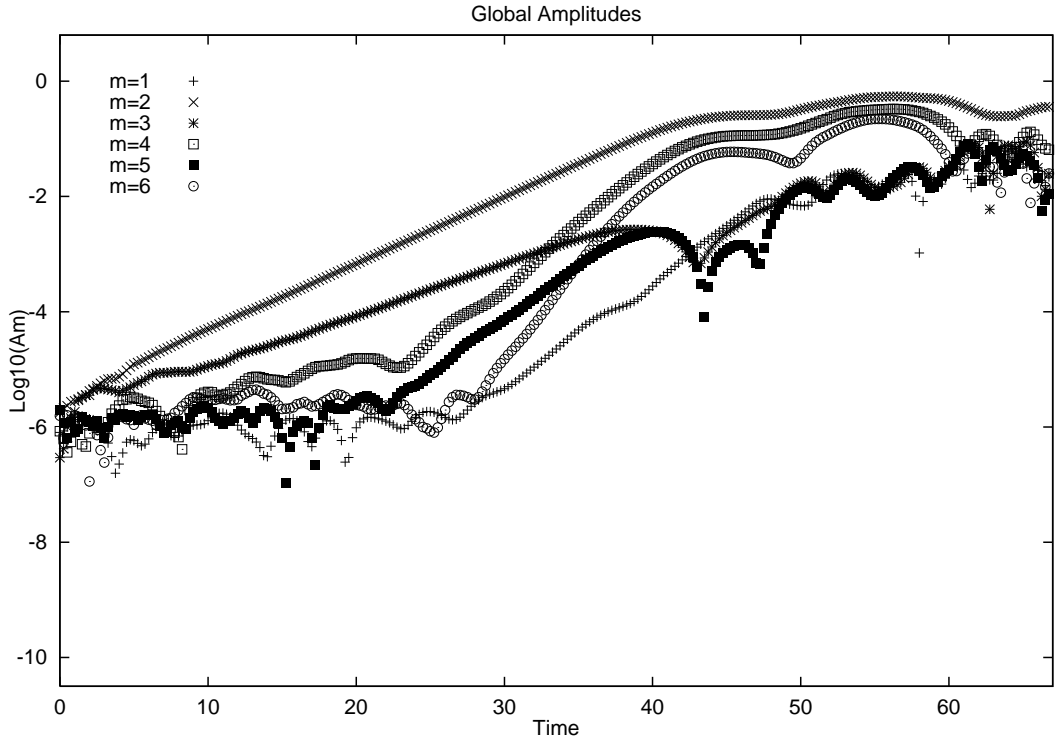}
\caption{The development of the global amplitudes
$\log_{10} A_m$ for $m = 1 - 6$ in the disk with equilibrium properties
shown in Figure 6.
During the linear phase, the $m=2$ and $m=3$ spirals are the most unstable
global modes. The exponential growth of the spiral modes saturates at
time $t \approx 40$.
$T_{unit}=1.34 \times 10^7$ years}
\end{figure}

\clearpage
\begin{figure}
\plotone{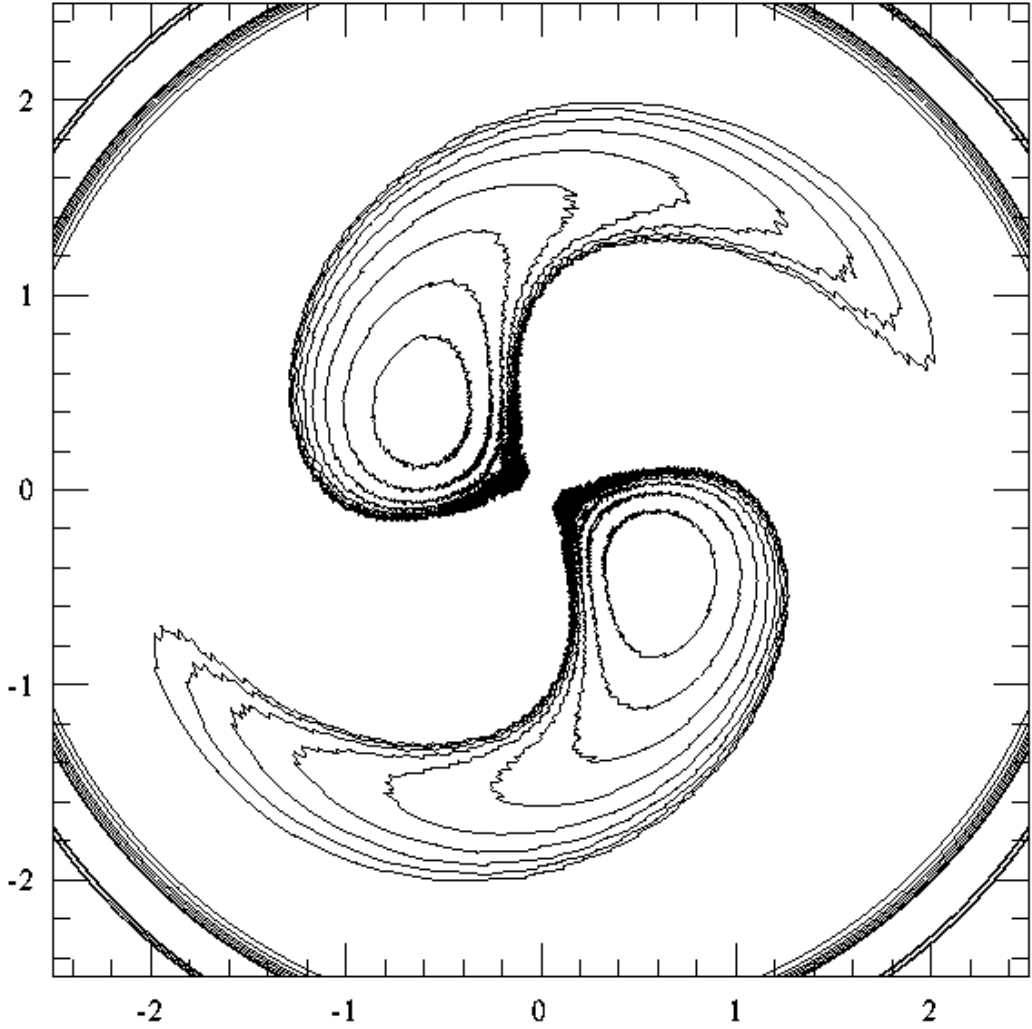}
\caption{Snapshot at $t=30$ of the surface density
perturbations in the unstable disk with $c_z/c_r = 1.0$.
Contour levels are logarithmically spaced between the maximum
value of density perturbations and one-hundredth of the maximum
density perturbations.
A two-armed open spiral pattern similar to the $m=2$ global mode found
in the linear modal analysis (Figure 7) emerges from the random
perturbations. Distance is measured in units $L_{unit}=2$ kpc.}
\end{figure}

\clearpage
\begin{figure}
\plotone{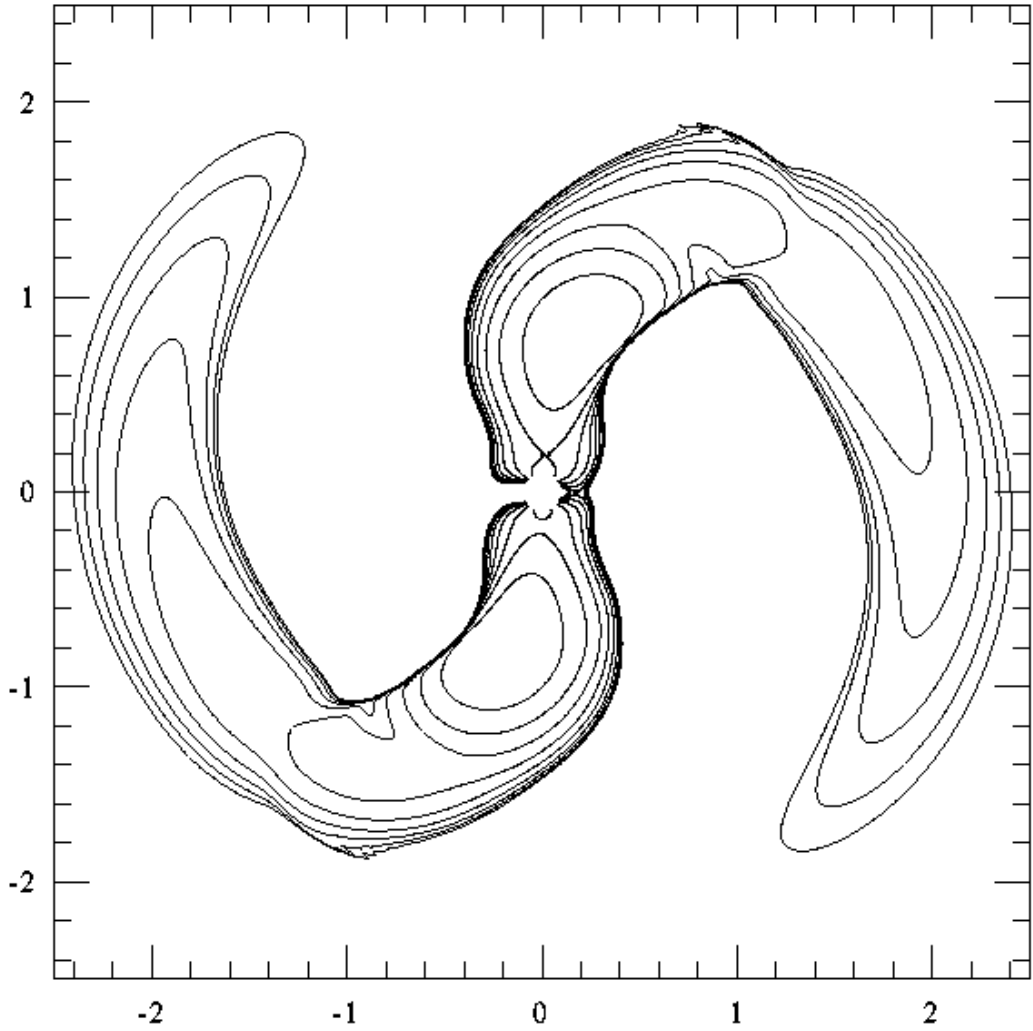}
\caption{Contour plots of the surface density
perturbations taken during the nonlinear stage of instability ($t=45$).
Shock fronts are formed at the concave edge of the spiral arms.
Contour levels are the same as in Figure 16. Distance is
in units $L_{unit}=2$ kpc.}
\end{figure}

\clearpage
\begin{figure}
\plotone{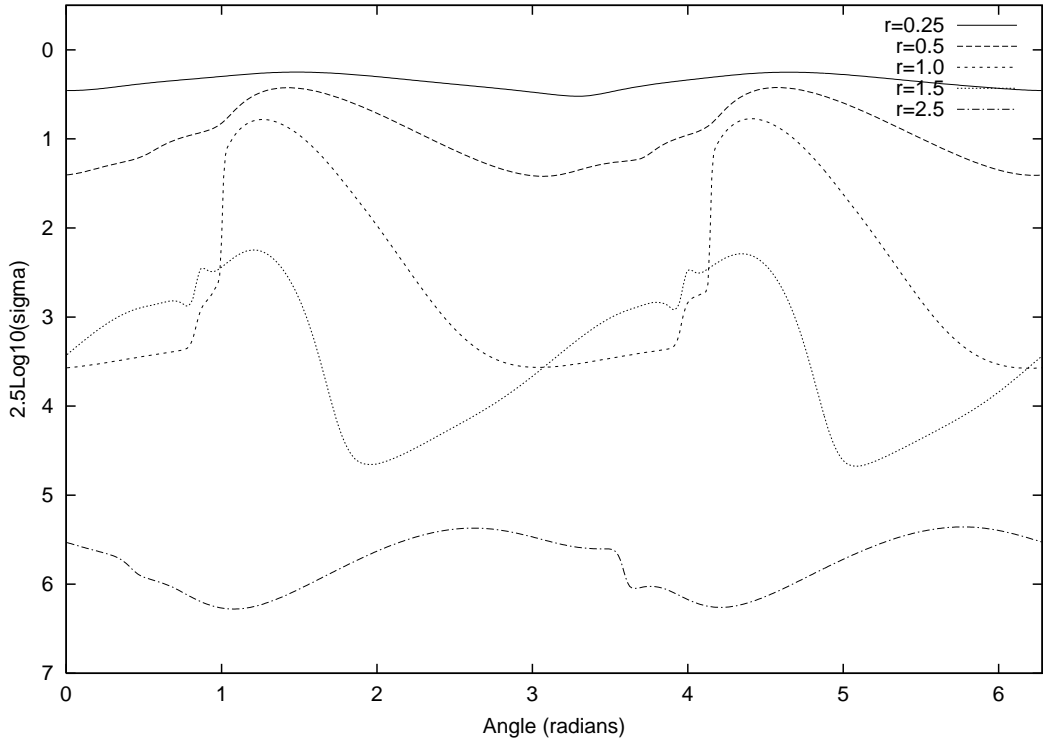}
\caption{Azimuthal variations of the ``surface brightness''
defined as $ -2.5 \log_{10} \sigma(t,R_N,\phi)$.
The azimuthal variations are computed at time $t=45$ in the five rings
with radii $R_1 =0.25$, $R_2 =0.5$,
$R_3 =1.0$,  $R_4 = 1.5$,
and $R_5 = 2.5$.
The theoretical azimuthal surface brightness variations increase with
radius for the central ring, and tend to decrease in outer rings.}
\end{figure}

\clearpage
\begin{figure}
\plotone{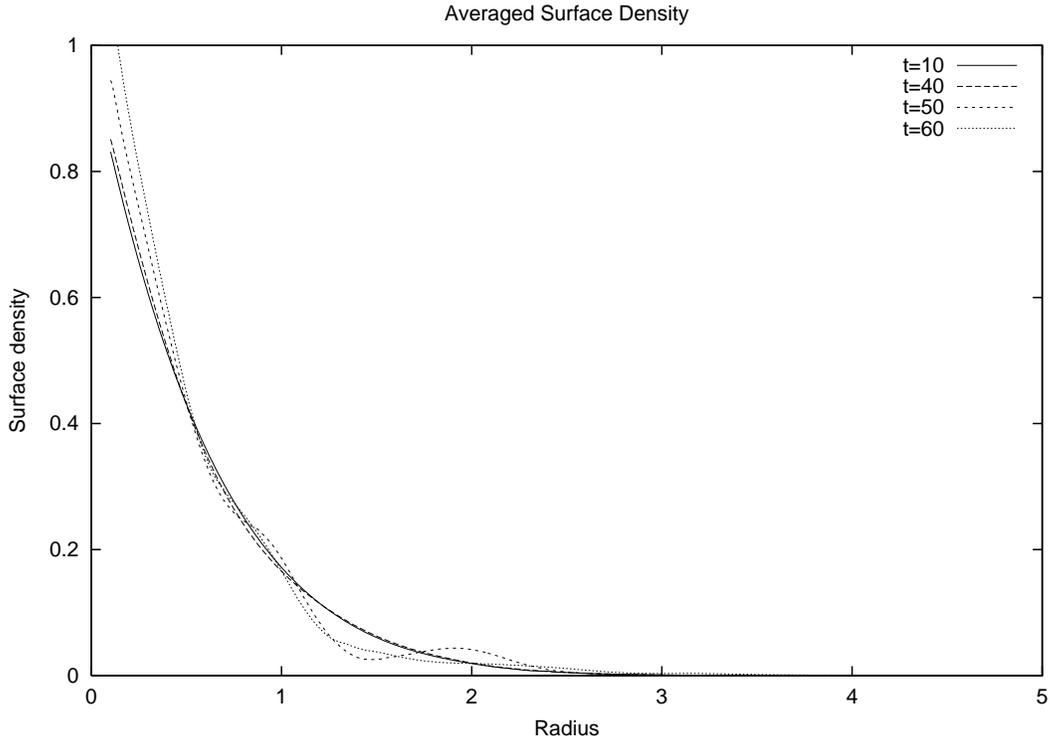}
\caption{Evolution of the surface density profile during the
extended dynamic range.
The nonlinear saturation phase in the development of the spiral instability
is accompanied by a steepening of the averaged surface density profile 
as compared to the observed surface density distribution in the disk of
NGC 1566. Radius is in units $L_{unit}=2$ kpc, surface density
is given in units of $2.5\times 10^9 M_{\odot}/kpc^2$.} 
\end{figure}

\clearpage
\begin{figure}
\plotone{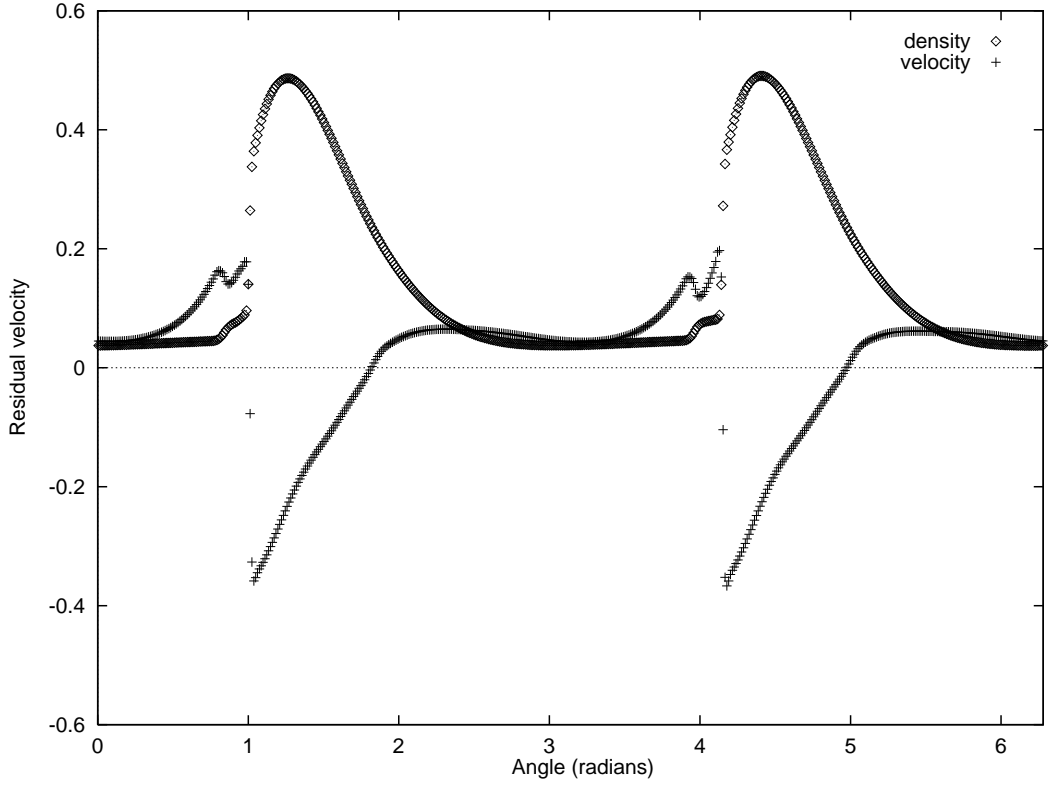}
\caption{The azimuthal profile of the residual velocity
(pluses) superimposed onto the surface density profile (diamonds)taken at time
$t=45$ at radius $R = 1.0$.
The residual velocity has strong gradient at the edge of the density
enhancement associated with the spiral shock.  Residual velocity
is given in units $V_{unit} = 149$ km/s, surface density is in units
$2.5\times 10^9 M_{\odot}/kpc^2$.}
\end{figure}

\newpage
\begin{table}
\caption{\label{table1}}
\vskip 1.5cm
\begin{tabular}{llll}
\hline \hline
Inner disk radius & $R_{in} = 0.1$ \\
Outer disk radius & $R_{out} = 5.0$ \\ 
Disk thickness & $z_0 = 0.35$\\
Surface brightness scalelength & $h_{SB} = 0.65$\\
\hline
``low'' curve: & $V_1 = 2.25~~ V_2 = 0.353~~ R_1=2.42~~ R_2=0.49$\\
``medium'' curve: & $V_1 = 2.678~~ V_2 = 0.549~~ R_1=2.18~~ R_2=0.32$\\
``upper'' curve: & $V_1 = 2.65~~ V_2 = 0.77~~ R_1=1.79~~ R_2=0.30$\\
\hline
One-component model: \\
Disk mass: & $ M_d = 1.78$\\
Polytropic index: & $\gamma = 2.0$\\
Polytropic coefficient & $K_s = 0.192$\\
\hline
Multi-component model:\\
Initial mass of gas & $ M_{g}=1.78$  \\
Gas polytropic index & $\gamma_g = 1.67$ \\
Gas polytropic coefficient & $K_g = 0.139$ \\
Initial mass of stars & $ M_{s}=0.0001$  \\
``Stellar''polytropic index & $\gamma_s = 2.0$ \\
``Stellar''polytropic coefficient & $K_s = 0.556$ \\
Initial mass of ``remnants''& $ M_{r}=0.0001$  \\
``Remnants''polytropic index & $\gamma_r = 2.0$ \\
``Remnants''popytropic coefficient & $K_r = 0.556$ \\
Stellar lifetime & $\tau = 1.0$ \\
Fraction of massive stars & $ \zeta = 0.12$ \\
Fraction of mass returned to ISM & $\eta = 0.9$ \\
SFR parameter & $C_2 = 0.1$ \\
\hline \hline
\end{tabular}
\end{table}

\end{document}